\documentclass[superscriptaddress,nofootinbib,notitlepage,preprintnumbers]{revtex4}
\pdfoutput=1
     
\usepackage{graphicx}				
\usepackage{amssymb}
\usepackage{amsmath, amsthm, amssymb}
\usepackage{slashed}
\usepackage{tikz}
\usepackage{rotating}
\usepackage{hyperref}
\usepackage{array}
\usepackage{color}
\usepackage{multirow}
\usepackage[utf8]{inputenc}

\begin{document}

\preprint{DESY 18-210}

\title{Simplified Models of Flavourful Leptoquarks}

\author{Ivo de Medeiros Varzielas}
\email{ivo.de@udo.edu}
\affiliation{CFTP, Departamento de F\'{i}sica, Instituto Superior T\'{e}cnico, Universidade de Lisboa, Avenida Rovisco Pais 1, 1049 Lisboa, Portugal}
\author{Jim Talbert} 
\email{james.talbert@desy.de}
\affiliation{Theory Group, Deutsches Elektronen-Synchrotron (DESY), Notkestra{\ss}e 85, 22607 Hamburg, Germany}

\begin{abstract}
We study the implications of single leptoquark extensions of the Standard Model (SM) under the assumption that their enhanced Yukawa sectors are invariant under global Abelian flavour symmetries already present in SM mass terms.  Such symmetries, assumed to be the `residual' subgroups of an ultra-violet flavour theory, have previously been considered in order to predict fermionic mixing angles.  Here we focus instead on their effect on the novel flavour structures sourced by the leptoquark representations that address
the present ${R}_{K^{(\star)}}$ anomalies in semileptonic rare $B$-decays.
Combined with existing flavour data, the residual symmetries prove to be extremely constraining; we find that the (quark-lepton) leptoquark Yukawa couplings fall within $\mathcal{O}(10)$ highly predictive patterns, each with only a single free  parameter, when `normal' (SM-like) hierarchies are assumed.
In addition, proton decay for the scalar SU(2) triplet representation is naturally avoided in the residual symmetry approach without relying on further model building.  Our results indicate that a simultaneous explanation for the ${R}_{K^{(\star)}}$ anomalies and the  flavour puzzle may be achieved in a simplified, model-independent formalism.
\end{abstract}

\maketitle

\section{Introduction}

Present data \cite{Aaij:2014ora,Aaij:2017vbb}  hint at lepton non-universality (LNU) and the breakdown of the Standard Model (SM) in the decay signatures of semileptonic rare $B$-decays.  In particular, the ratio observables 
\begin{align}
\label{eq:RK}
{R}_{K^{(\star)}, [a,b]} &= \frac{\int_{a}^{b} \, dq^{2} \, \left[d\Gamma (B \rightarrow K^{(\star)} \mu^{+}\mu^{-})/ dq^{2} \right]}{\int_{a}^{b} \, dq^{2} \, \left[d\Gamma (B \rightarrow K^{(\star)} e^{+}e^{-})/ dq^{2} \right]} \, , 
\end{align}
with $q^{2}$ the invariant di-lepton mass and $\left[a,b\right]$ representing bin boundaries (in GeV$^{2}$), are currently each measured at 2-3$\sigma$ deviations away from their SM expectations (see e.g. \cite{Hiller:2003js,Bordone:2016gaq}),
as seen in Table \ref{tab:experiment}.
These observables are particularly interesting because hadronic theory uncertainties are cancelled by virtue of the ratio definition \cite{Hiller:2003js}, and hence ${R}_{K^{(\star)}}$ are  clean tests of the SM

\begin{table} [t]
\centering
{\renewcommand{\arraystretch}{1.5}
\begin{tabular}{|c||c|c|c|}
\hline
Ratio & Bin (GeV$^2$) & Data & Experimental Reference \\
\hline
\hline
${R}_{K}$ & [1, 6] & $0.745^{+0.090}_{-0.074} \pm 0.036$ & \text{LHCb}\,\,\cite{Aaij:2014ora} \\
\hline
 \multirow{2}{*}{${R}_{K^\star}$} & [1.1, 6.0] & $0.685^{+0.113}_{-0.069} \pm 0.047$ & \text{LHCb}\,\,\cite{Aaij:2017vbb} \\
\cline{2-4}
& [0.045, 1.1] &  $0.66^{+0.11}_{-0.07} \pm 0.03$ & \text{LHCb}\,\,\cite{Aaij:2017vbb} \\
\hline
\end{tabular}}
\caption{Data on  ${R}_{K^{(\star)}}$ by  the LHCb collaboration. 
}
\label{tab:experiment}
\end{table}

The potentially anomalous data in Table \ref{tab:experiment} has prompted a flurry of theoretical and phenomenological studies over the last few years.  From a model-independent perspective, global fits to effective field theory (EFT) operators \cite{Hiller:2014yaa,Capdevila:2017bsm,Altmannshofer:2017yso,DAmico:2017mtc,Hiller:2017bzc,Ciuchini:2017mik,Alok:2017sui} have concluded that new physics contributions to four-fermion contact interactions mediated by left-handed (LH) quark currents, i.e. to combinations of the $C^{l}_{9}$ and/or $C^{l}_{10}$ (with $l = e, \mu$) Wilson coefficients of the weak effective Hamiltonian (see e.g. \cite{Hiller:2017bzc} for a definition of these coefficients), are sufficient to explain observations.  On the other hand, numerous model-specific explanations for some or all of the data have also been offered, including $Z^{\prime}$, flavour symmetric, leptoquark, composite- and multi- higgs approaches \cite{Gripaios:2014tna,Varzielas:2015iva,Gripaios:2015gra,Bauer:2015knc,Arnan:2016cpy,Hiller:2016kry,Crivellin:2016ejn,Crivellin:2017zlb,Alonso:2017bff,Bonilla:2017lsq,King:2017anf,Aloni:2017ixa,Assad:2017iib,Calibbi:2017qbu, Hiller:2018wbv,deMedeirosVarzielas:2018bcy,Grinstein:2018fgb, Fornal:2018dqn}.  Some of these models, e.g. \cite{Varzielas:2015iva,Crivellin:2016ejn,deMedeirosVarzielas:2018bcy,Grinstein:2018fgb}, also address the SM flavour puzzle, the unexplained quantizations of the 20-22 free parameters associated to fermionic mass and mixing.  In this paper we also explore the simultaneous explanation of $\mathcal{R}_{K^{(\star)}}$ with the flavour puzzle via an analysis of the global flavour symmetries of the SM Yukawa sector when enhanced by a single leptoquark field.  In this way we strike an intermediate path between a fully model-independent EFT analysis and an explicit model of new physics.

To better motivate our approach, let us first consider the SM in its unbroken phase where, absent the Yukawa couplings, it exhibits a $U(3)^{5}$ global flavour symmetry \cite{Chivukula:1987py}, with one $U(3)$ rotational invariance associated to each of the chiral fermionic sectors.  This symmetry is accidental --- a priori, no gauge structure nor dynamical content is associated to it.  However, it may very well hint at an underlying mechanism controlling flavour.  Indeed, a popular approach to studying flavour is via the principle of `Minimal Flavour Violation' (MFV) \cite{DAmbrosio:2002vsn}, in which one assumes that the only $U(3)^{5}$ flavour violating terms are the Yukawa interactions, whose couplings are then promoted to spurions with (symmetry-breaking) background expectation values.  One then proceeds to build an effective theory out of higher dimensional operators generated via successive spurion insertions.  MFV provides a roughly model-independent framework for exploring the dynamics of flavour and its scale in collider processes.  One the other hand, it offers no explanation for the flavour puzzle, and its extension to the lepton sector is not unique \cite{Cirigliano:2005ck}. Unfortunately, one also finds that experimental data (e.g. $b \rightarrow s\, \mu^{+}\mu^{-}$, $b \rightarrow s \,\tau^{+}\tau^{-}$)  exclude an extension of the MFV hypothesis to leptoquark models explaining Table \ref{tab:experiment}, at least in the linear regime \cite{Aloni:2017ixa}. 

Intriguingly, the SM Yukawa sector also exhibits accidental symmetries in its \emph{broken} phase, i.e. after the Higgs has obtained its vacuum expectation value (vev) during electroweak symmetry breaking (EWSB), such that the SM fermions (except perhaps the neutrinos) are rendered massive. 
To see this, let us follow earlier discussions \cite{Lam:2007qc,Hernandez:2012ra} and consider the SM leptonic mass sector, assuming a Majorana neutrino term generated (e.g.) with a type-I seesaw mechanism \cite{Minkowski:1977sc}:
\begin{equation}
\label{eq:SMLyuk}
\mathcal{L}_{\text{mass}} = 
\bar{l}_{L}\, m_{l}\,E_{R} + \frac{1}{2} \bar{\nu}^{c}_{L}\, m_{\nu} \, \nu_{L} +  \text{...} + \text{h.c.}
\end{equation}
Here $l_L$ and $\nu_L$ are the charged lepton and neutrino components of the leptonic $SU(2)$ doublets and $E_R$ is the $SU(2)$ singlet.
$m_{l,\nu}$ denote diagonal mass matrices. By examining \eqref{eq:SMLyuk}, one notes that the Majorana neutrino mass term is naturally invariant under a Klein $Z_{2} \times Z_{2}$ transformation of the neutrinos:
\begin{equation}
\label{eq:klein}
\nu_{L} \rightarrow T_{\nu i}\, \nu_{L}, \,\,\,\,\,\,\,\,\,\, m_{\nu} \rightarrow T_{\nu i}^{T}\, m_{\nu}\, T_{\nu i} = m_{\nu}   \, , 
\end{equation}
where the $Z_{2}$ generators $T_{\nu i}$ can be generically written as
\begin{equation}
\label{eq:klein2}
T_{\nu 1} = diag \left(1,-1,-1 \right), \,\,\,\,\,\,\,\,\,\,T_{\nu 2} = diag \left(-1,1,-1 \right) \, . 
\end{equation}
On the other hand, the charged lepton mass term is subject to a $U(1)^{3}$ symmetry associated to independent rephasings of each generation. The action of this symmetry can be represented by a generator $T_{l}$:
\begin{equation}
\label{eq:CLreps}
l_{L} \rightarrow T_{l} \, l_{L}, \,\,\,\,\,\,\,\,\,\, E_{R} \rightarrow T_{l}\,E_{R},\,\,\,\,\,\,\,\,\,\,T_{l} = diag \left(e^ {i \alpha_{l}},e^{i \beta_{l}},e^{ i \gamma_{l}} \right)  \, . 
\end{equation} 
Analogous $U(1)^{3}$ symmetries, with associated generator representations $T_{u,d}$, also exist in the quark sector and, if they are instead Dirac particles, the neutrino sector.  Furthermore, while the invariance of \eqref{eq:SMLyuk} (and its quark analogue) under $T_{u,d,l,\nu}$ is shown in the mass basis, it of course also rotates to the flavour basis where information regarding fermionic mixing can be extracted, and so observed patterns may be understood with the residual symmetry mechanism \cite{Lam:2007qc,Hernandez:2012ra,deAdelhartToorop:2011re, Lam:2012ga, Holthausen:2012wt, Holthausen:2013vba, King:2013vna, Lavoura:2014kwa, Joshipura:2014pqa, Joshipura:2014qaa, Talbert:2014bda, Yao:2015dwa, King:2016pgv, Varzielas:2016zuo,Yao:2016zev}.\footnote{This type of Abelian phase symmetry was also studied in the context of multi-Higgs-doublet models in \cite{Ferreira:2010ir,Serodio:2013gka,Ivanov:2013bka}.}

The SM mass sector is therefore left invariant under the actions of \emph{residual symmetries} generated by $T_{u,d,l,\nu}$, which can be interpreted as the generators of residual subgroups $\mathcal{G}_{u,d,l,\nu}$ of a parent flavour symmetry $\mathcal{G}_{\mathcal{F}}$.  For example, an illustrative breaking chain from the ultra-violet (UV) might go as
\begin{equation}
\label{eq:GF}
\mathcal{G_{F}}  \rightarrow \begin{cases}
				\mathcal{G_{L}}   \rightarrow \begin{cases} 
										\mathcal{G_{\nu}}
										\\
										\mathcal{G_{\text{l}}}
										\end{cases} \\
				\mathcal{G_{Q}} \rightarrow \begin{cases}
										\mathcal{G_{\text{u}}}
										\\
										\mathcal{G_{\text{d}}}
										\end{cases}
				\end{cases}
\end{equation}
It is important to emphasize that, in identifying the residual subgroups $\mathcal{G}_{u,d,l,\nu}$ in \eqref{eq:SMLyuk}, we are of course not implying that SU(2)$_{L}$ is broken before EWSB.  Instead, the residual symmetries distinguish members of LH doublets only after EWSB, when the Higgs vev couples differently to $u_{R}$ and $d_{R}$ fields.  The phases of the $T$ generators amount to signatures of the UV parent symmetry/theory which commutes entirely with the SM gauge group, including SU(2)$_{L}$. The distinct action on members of LH doublets can be assumed to originate in the initial breaking of $\mathcal{G}_{\mathcal{F}}$ (or $\mathcal{G}_{\mathcal{L},\mathcal{Q}}$), perhaps via flavon fields acquiring vevs in specific directions of flavour space.  Different breaking directions in each fermion sector will then lead to different $\mathcal{G}_{u,d,l,\nu}$.   This also explains why the SM weak interactions do \emph{not} respect $\mathcal{G}_{u,d,l,\nu}$ --- one only expects them to hold in Yukawa(-like) terms that, in order to respect $\mathcal{G_{F}}$, must be enhanced by flavon field insertions, and can therefore be considered effective terms in an operator product expansion (or which exhibit some other mechanism for breaking $\mathcal{G_{F}}$).  As it turns out, the residual symmetry mechanism generalizes the symmetry breaking of entire classes of flavour models --- for pedagogical reviews of such models (and where the compatibility with SU(2)$_{L}$ can be seen), see \cite{King:2013eh,Altarelli:2010gt}.  For an explicit, complete model that realizes residual symmetries see (e.g.) the original construction in \cite{Altarelli:2005yx} and its UV completion \cite{Varzielas:2010mp}.\footnote{To account for the observed reactor angle it is possible to modify the original model and its respective UV completion as in \cite{Varzielas:2012ai}, or to use the semi-direct approach predicting only one column of the mixing matrix as in \cite{Varzielas:2012pa}.}  

\begin{table} [t]
\centering
{\renewcommand{\arraystretch}{1.5}
\begin{tabular}{|c||c|c|c|}
\hline
Leptoquark & Representation & $C_9$ and $C_{10}$ Relation & $\mathcal{R}_{K^{(\star)}}$ \\
\hline
\hline
$\Delta_{3}$ & $\left(\bar{3}, 3, 1/3\right)$ & $C_9 = - C_{10}$ & $\mathcal{R}_{K} \simeq \mathcal{R}_{K^{\star}} < 1$\\ 
\hline
 \multirow{2}{*}{$\Delta_{1}^{\mu}$} &  \multirow{2}{*}{$\left(3, 1, 2/3\right)$} & $C_9 = - C_{10}$ & $\mathcal{R}_{K} \simeq \mathcal{R}_{K^{\star}} < 1$\\ 
 \cline{3-4}
 & & $C_{9} = C_{10}$ & $\mathcal{R}_{K} \simeq \mathcal{R}_{K^{\star}} \simeq 1$ \\
\hline
$\Delta_{3}^{\mu}$ & $\left(3, 3, 2/3\right)$ & $C_9 = - C_{10}$ & $\mathcal{R}_{K} \simeq \mathcal{R}_{K^{\star}} < 1$\\ 
\hline
\end{tabular}}
\caption{Relationships implied for the Wilson coefficients $C_{9,10}$, and the corresponding predictions for $\mathcal{R}_{K^{(\star)}}$, for the three leptoquarks we consider. } 
\label{tab:LQs}
\end{table}

In what follows, we apply the same analysis in \eqref{eq:SMLyuk}-\eqref{eq:CLreps} to the SM appended by a single leptoquark\footnote{The physics of leptoquarks is thoroughly reviewed in \cite{Dorsner:2016wpm}, and we follow their charge normalizations here as well.} sourcing tree-level couplings between quarks and leptons, with the aim of understanding the experimental observations in Table \ref{tab:experiment}.  
In particular, we study a scalar leptoquark transforming as a triplet of $SU(2)$ (referred to as $S_3$ in \cite{Hiller:2017bzc}), and two vector leptoquarks transforming as either a singlet or triplet of $SU(2)$ (referred to as $V_1$ and $V_{3}$ respectively in \cite{Hiller:2017bzc}).  All three are colour triplets and give excellent fits to the data.
The full representations of these fields under the SM as well as the relationship they imply between the $C_{9}$ and $C_{10}$ Wilson coefficients, and ultimately $\mathcal{R}_{K^{(\star)}}$, is given in Table \ref{tab:LQs} (taken from \cite{Hiller:2017bzc}).  Of course, leptoquark extensions of the SM have been studied in light of Table \ref{tab:experiment} while also considering the flavour problem before  \cite{Varzielas:2015iva,Crivellin:2016ejn,deMedeirosVarzielas:2018bcy,Grinstein:2018fgb}, albeit with different assumptions.

Outside of this enhanced field content, the core assumption of our study is that, regardless of the origins and structure of $G_{\mathcal{F}}$, its scale, or the mechanism associated to its breaking, it does so to the residual symmetries present in \eqref{eq:GF}, and furthermore that these symmetries also leave the new leptoquark Yukawa couplings invariant.  That is, we promote the accidental actions of $T_{u,d,l,\nu}$ to those of physical symmetries, which we use to define a simplified model space whose phenomenology can be studied without reference to UV dynamics.  We will show that the consequences of this construction are extremely constraining.

The paper develops as follows:  in Section \ref{sec:LQ} we review the enhanced Yukawa sector upon including Table \ref{tab:LQs} into our field content.  We then discuss the application of residual symmetries in the full Yukawa Lagrangian, and show that for the charged state providing tree-level BSM contributions to \eqref{eq:RK}, only a handful of Yukawa patterns are permitted.  In this section we also present the current experimental bounds on the relevant coupling matrix, and parameterize the combined data into a form that is inspired by SM Yukawa hierarchies.  We also briefly discuss the implied hadron collider phenomenology.  Then, in Section \ref{sec:otherstates}, we derive the further constraints implied when all of the charged states sourced after isospin decomposition are included for the scalar triplet, ultimately finding that there are only nine unique patterns allowed in the quark-lepton sector. We further show that additional relationships in the symmetry generators of the up and down sectors are sufficient to avoid proton decay.  We then extend our analysis to the vector triplet and singlet scenarios in Section \ref{sec:otherstatesV1V3}, where the same patterns of couplings emerge. Finally, before concluding in Section \ref{sec:conclude}, in Section \ref{sec:reduce} we briefly comment on how our conclusions would change were we to allow for a reduced symmetry at the level of the SM Lagrangian.  Additionally, we collect all of the explicit matrices derived in Section \ref{sec:otherstates} in Appendix \ref{sec:A} for easy reference.

\section{Leptoquark Yukawa Couplings and Residual Symmetries}
\label{sec:LQ}

There are 12 potential Yukawa couplings for leptoquarks charged under the SM gauge symmetries, not all of which are relevant for addressing the $\mathcal{R}_{K^{(\star)}}$ anomalies.  They are categorized, including their effective vertices, in \cite{Dorsner:2016wpm,Hiller:2016kry}.  Importantly, and unlike in the SM, there are potentially physical Yukawa couplings with right-handed field rotations, and hence we initially assume that all fermion fields undergo some sort of transformation, similar to \cite{Hiller:2016kry}:
\begin{align}
\nonumber
u_{L} &\rightarrow U_{u} u_{L}\,, \,\,\,\,\,\,\,\,\,\,\,\,\,\,\,\,\,\,\,\,\, d_{L} \rightarrow U_{d} d_{L}\,, \,\,\,\,\,\,\,\,\,\,\,\,\,\,\,\,\,\,\,\,\, l_{L}\rightarrow U_{l} l_{L}\,, \,\,\,\,\,\,\,\,\,\,\,\,\,\,\,\,\,\,\,\,\, \nu_{L} \rightarrow U_{\nu} \nu_{L} \,,  \\
\label{eq:rotate2}
u_{R} &\rightarrow U_{U} u_{R}\,,\,\,\,\,\,\,\,\,\,\,\,\,\,\,\,\,\,\,\,\, d_{R} \rightarrow U_{D} d_{R}\,, \,\,\,\,\,\,\,\,\,\,\,\,\,\,\,\, E_{R} \rightarrow U_{E} E_{R}\,,\,\,\,\,\,\,\,\,\,\,\,\,\,\,\,\,\,\,  \nu_{R} \rightarrow U_{R} \nu_{R} \, , 
\end{align}
such that leptoquark Yukawas transform under a basis rotation as
\begin{equation}
Y_{AB} \rightarrow U^{(T,\dagger)}_{A} Y_{AB} U_{B}   \, , 
\end{equation}
with $A,B$ arbitrary quark and lepton fields and where the relevant operation on $U^{(T,\dagger)}_A$ is determined by the conjugation structure of $A$ and $B$.

As mentioned above, in this paper we study the three phenomenologically interesting leptoquarks of Table \ref{tab:LQs}.  Written explicitly in $SU(2)$ space, the Yukawa interactions of these fields go as
\begin{align}
\nonumber
\Delta_{3}&: \,\,\,\,\,&&\mathcal{L} \supset y_{3, ij}^{LL} \bar{Q}_{L}^{C\,i,a} \epsilon^{ab} (\tau^{k} \Delta_{3}^{k})^{bc} L_{L}^{j,c} + z_{3,ij}^{LL} \bar{Q}_{L}^{C\,i,a}\epsilon^{ab}((\tau^{k}\Delta_{3}^{k})^{\dagger})^{bc}Q_{L}^{j,c} + \text{h.c.} \\
\nonumber
\Delta_{1}^{\mu}&: \,\,\,\,\,&&\mathcal{L} \supset x_{1,ij}^{LL} \bar{Q}_{L}^{i,a} \gamma^{\mu} \Delta_{1,\mu} L_{L}^{j,a} + x_{1,ij}^{RR} \bar{d}^{i}_{R} \gamma^{\mu} \Delta_{1,\mu} e_{R}^{j} + x_{1,ij}^{\overline{RR}} \bar{u}_{R}^{i} \gamma^{\mu} \Delta_{1,\mu} \nu_{R}^{j} + \text{h.c.}\\
\label{eq:LLyukSU2}
\Delta_{3}^{\mu}&: \,\,\,\,\,&&\mathcal{L}  \supset  x_{3,ij}^{LL} \bar{Q}_{L}^{i,a} \gamma^{\mu} \left(\tau^{k} \Delta_{3,\mu}^{k}\right)^{ab} L_{L}^{j,b}  + \text{h.c.} 
\end{align}
Here $\lbrace i,j \rbrace$ are flavour indices, $\lbrace a,b \rbrace$ are $SU(2)$ indices, and $k = 1,2,3$ for the Pauli matrices.  Colour indices are left implicit.  The $y^{LL}$ and $x^{LL}$ clearly source tree-level couplings between leptons and quarks, and so are relevant to our study of the $\mathcal{R}$ anomalies. Following \cite{Dorsner:2016wpm}, we define new combinations of the components of $\Delta_{3}^{(\mu)}$ given by
\begin{align}
\nonumber
&\Delta_{3}^{4/3} = \left(\Delta_{3}^{1} - i \Delta_{3}^{2} \right)/\sqrt{2}, \,\,\,\,\,\,\,\,\,\,\,
&&\Delta_{3}^{-2/3} = \left(\Delta_{3}^{1} + i \Delta_{3}^{2} \right)/\sqrt{2}, \,\,\,\,\,\,\,\,\,\,\,
&&\Delta_{3}^{1/3} = \Delta_{3}^{3}  \, ,  \\
&\Delta_{3}^{\mu,5/3}= \left(\Delta_{3}^{\mu,1} - i \Delta_{3}^{\mu,2} \right)/\sqrt{2}, \,\,\,\,\,\,\,\,\,\,\,
&&\Delta_{3}^{\mu,-1/3} = \left(\Delta_{3}^{\mu,1} + i \Delta_{3}^{\mu,2} \right)/\sqrt{2}, \,\,\,\,\,\,\,\,\,\,\,
&&\Delta_{3}^{\mu,2/3} = \Delta_{3}^{\mu,3}  \, , 
\label{eq:combo1}
\end{align}
where on the right-hand side (RHS) superscripts denote SU(2) components of $\Delta$, and on the left-hand side (LHS) they denote the electric charges of the newly defined states.  Contracting the $SU(2)$ indices of \eqref{eq:LLyukSU2}, one obtains 
\begin{align}
\nonumber
\Delta_{3}&: &&\mathcal{L} \supset -(U_{d}^{T} y_{3}^{LL} U_{\nu})_{ij} \bar{d}^{C \, i}_{L} \Delta^{1/3}_{3} \nu_{L}^{j} - \sqrt{2} (U_{d}^{T} y_{3}^{LL} U_{l})_{ij} \bar{d}^{C \, i}_{L} \Delta^{4/3}_{3} l_{L}^{j}  \\
\nonumber
&&& + \sqrt{2} (U_{u}^{T} y_{3}^{LL} U_{\nu})_{ij} \bar{u}^{C \, i}_{L} \Delta^{-2/3}_{3} \nu_{L}^{j} -(U_{u}^{T} y_{3}^{LL} U_{l})_{ij}\bar{u}^{C \, i}_{L} \Delta^{1/3}_{3} l_{L}^{j}  \\
\nonumber
&&&+ \text{h.c.} \\
\nonumber
\Delta_{1}^{\mu}&: &&\mathcal{L} \supset (U_{u}^{\dagger} x_{1}^{LL} U_{\nu})_{ij} \bar{u}^{ i}_{L} \gamma^{\mu} \Delta_{1,\mu} \nu_{L}^{j} +  (U_{d}^{\dagger} x_{1}^{LL} U_{l})_{ij} \bar{d}^{i}_{L} \gamma^{\mu} \Delta_{1,\mu} l_{L}^{j}  \\
\nonumber
&&& + (U_{D}^{\dagger}x_{1}^{RR} U_{E})_{ij} \bar{d}^{i}_{R} \gamma^{\mu} \Delta_{1,\mu} E_{R}^{j} + (U_{U}^{\dagger} x_{1}^{RR} U_{R})_{ij}\bar{u}^{ i}_{R} \gamma^{\mu} \Delta_{1,\mu} \nu_{R}^{j}  \\
\nonumber
&&&+ \text{h.c.} \\
\nonumber
\Delta_{3}^{\mu}&: &&\mathcal{L} \supset -(U_{d}^{\dagger} x_{3}^{LL} U_{l})_{ij} \bar{d}^{ i}_{L} \gamma^{\mu} \Delta^{2/3}_{3,\mu} \l_{L}^{j} +  (U_{u}^{\dagger} x_{3}^{LL} U_{\nu})_{ij} \bar{u}^{ i}_{L} \gamma^{\mu} \Delta^{2/3}_{3,\mu} \nu_{L}^{j}  \\
\nonumber
&&& + \sqrt{2} (U_{d}^{\dagger} x_{3}^{LL} U_{\nu})_{ij} \bar{d}^{i}_{L} \gamma^{\mu} \Delta^{-1/3}_{3,\mu} \nu_{L}^{j} + \sqrt{2}(U_{u}^{\dagger} x_{3}^{LL} U_{l})_{ij}\bar{u}^{ i}_{L} \gamma^{\mu} \Delta^{5/3}_{3,\mu} l_{L}^{j}  \\
\label{eq:LLyukFLAV}
&&&+ \text{h.c.}
\end{align}
for the lepton-quark terms and
\begin{align}
\nonumber
\mathcal{L} \supset &-(U_{d}^{T} z_{3}^{LL} U_{u})_{ij} \bar{d}^{C \, i}_{L} \Delta^{1/3,\star}_{3} u_{L}^{j} - \sqrt{2} (U_{d}^{T} z_{3}^{LL} U_{d})_{ij} \bar{d}^{C \, i}_{L} \Delta^{-2/3,\star}_{3} d_{L}^{j}\\
\nonumber
& + \sqrt{2} (U_{u}^{T} z_{3}^{LL} U_{u})_{ij} \bar{u}^{C \, i}_{L} \Delta^{4/3,\star}_{3} u_{L}^{j} -(U_{u}^{T} z_{3}^{LL} U_{d})_{ij}\bar{u}^{C \, i}_{L} \Delta^{1/3,\star}_{3} d_{L}^{j} \\
\label{eq:LLyukFLAV2}
&+ \text{h.c.}
\end{align}
for the quark-quark coupling of $\Delta_{3}$.  In both \eqref{eq:LLyukFLAV} and \eqref{eq:LLyukFLAV2} we have changed bases via \eqref{eq:rotate2}.   

As is clear, the vector states $\Delta_{(1,3)}^{\mu}$ do not permit diquark operators sourcing proton decay, and  a careful examination reveals that for the scalar $\Delta_{3}$ the diquark operator is anti-symmetric under $SU(3)_C$, and so gauge invariance requires \cite{Davies:1990sc,Vecchi:2011ab}
\begin{equation}
\label{eq:antisymm}
z_{3,ij}^{LL} = - z_{3,ji}^{LL} \, , 
\end{equation}
which automatically forbids proton decay through the diagonal entries of the up-up and down-down operators of \eqref{eq:LLyukFLAV2}, even in the mass basis. These couplings can source proton decay via their off diagonal elements, as can the couplings in the up-down operators in \eqref{eq:LLyukFLAV2} (see e.g. the analysis of proton decay induced by leptoquarks in \cite{Nath:2006ut}). The dangerous matrix elements must be avoided in any explicit model.  This is often achieved with a new symmetry under which the leptoquark is charged non-trivially, giving neutral $\bar{Q}Q$, whereas other combinations need not be and the associated terms can remain invariant by canceling the charge of the leptoquark.  In what follows we do not make this model-dependent assumption a priori, but instead show in Section \ref{sec:QQ} that the dangerous terms of \eqref{eq:LLyukFLAV2} can be killed with simple phase relationships in addition to those derived upon application of the residual symmetry principle in \eqref{eq:LLyukFLAV}-\eqref{eq:LLyukFLAV2}.\footnote{Indeed, in the discussion that follows our modus operandi will be to analyze the symmetry and experimental consequences sourced from \eqref{eq:LLyukFLAV}, the terms generating the phenomenological signatures of interest, and to then return to \eqref{eq:LLyukFLAV2} to study their implications in the diquark sector.}  In other words, the residual symmetry mechanism provides adequate protection against proton decay without the need for additional model building.

From \eqref{eq:LLyukFLAV}-\eqref{eq:LLyukFLAV2} we see a host of structures similar to the terms in \eqref{eq:SMLyuk}.  If we assume that the  leptoquark Yukawas are invariant under a residual symmetry transformation generated by $X$, then up to constant prefactors the $(...)_{ij} \equiv \lambda_{(QL,QQ)}$ terms are analogous to basis-transformed mass matrices which must be invariant under transformations of the form
\begin{equation}
 \lambda_{(QL,QQ)}\rightarrow X^{(T,\dagger)}_{V_{1}}  \lambda_{(QL,QQ)} X_{V_{2}} \overset{!}{=} \lambda_{(QL,QQ)}   \, , 
\end{equation}
where $V_{1}$ and $V_{2}$ represent arbitrary rotations depending on the terms in \eqref{eq:LLyukFLAV}, and $X_{V_{1,2}}$ clearly depend on the basis of $\lambda_{(QL,QQ)}$.
However, the new couplings in \eqref{eq:LLyukFLAV} connect SM leptons to SM quarks!  We must take care then to understand exactly how a parent flavour symmetry, upon breaking to residuals in some or all of these sectors simultaneously, is actioned in bases relevant to understanding observed experimental signatures.  

Let us focus for the moment on the scalar leptoquark Yukawa term coupling down quarks to charged leptons, as it can source tree-level contributions to $\mathcal{R}_{K^{(\star)}}$  observables.  Removing flavour indices for simplicity, the Yukawa sector includes the following terms
\begin{equation}
\label{eq:LQYUK1}
\mathcal{L}  \supset  \bar{l}_{L}\, m_{l}\,E_{R}   +  \bar{d}_{L}\, m_{d} \, d_{R} + \bar{d}^{C}_{L} \,\lambda_{dl}\, l_{L} \, \Delta^{4/3}_{3} + \text{h.c.}
\end{equation}
where we have chosen to work in the mass basis of the down quarks and charged leptons, giving diagonal $m_{l,d}$.  As a result, the leptoquark Yukawa coupling is generically non-diagonal, and we can identify its rows and columns in a generation specific way \cite{Varzielas:2015iva}:    
\begin{equation}
\label{eq:genericyuk}
-\sqrt{2}\,\left(U_{d}^{T}\,y_{3}^{LL}\,U_{l} \right) \equiv \lambda_{dl} =
\left(
\begin{array}{ccc}
\lambda_{de} & \lambda_{d\mu} & \lambda_{d\tau}  \\
\lambda_{se}  & \lambda_{s\mu}  & \lambda_{s\tau}     \\
\lambda_{be}  & \lambda_{b\mu}   & \lambda_{b\tau}  
\end{array}
\right) \, . 
\end{equation}
We now make our core assumption, namely that the residual symmetries of the SM mass terms also hold in the leptoquark Yukawa terms. This assumption can be implemented naturally in models where the same flavons give rise to the different types of Yukawa (see e.g. the flavon models in \cite{Varzielas:2015iva}). We therefore apply the residual transforms
\begin{align}
\label{eq:Xtransform}
d_{L,R} \rightarrow T_{d} \,d_{L,R}, \,\,\,\,\,\,\,\,\,\, l_{L} \rightarrow T_{l} \,l_{L},\,\,\,\,\,\,\,\,\,\,E_{R} \rightarrow T_{l}\, E_{R}  \, , 
\end{align}
where the residual generators $T_{l,d}$ are generically represented by diagonal matrices of arbitrary phases, $T_{j \in l,d} = \text{diag}\left(e^{i \alpha_{j}}, e^{i\beta_{j}}, e^{i \gamma_{j}} \right)$, as in \eqref{eq:CLreps}.   We observe that the corresponding residual symmetry constraint on the leptoquark term of \eqref{eq:LQYUK1} is given by 
 \begin{equation}
\label{eq:LQoverconstrain}
    \left(
\begin{array}{ccc}
e^{i(\alpha_{d}+\alpha_{l})}\,\lambda_{de} & e^{i(\alpha_{d}+\beta_{l})}\, \lambda_{d\mu} &  e^{i(\alpha_{d}+\gamma_{l})}\,\lambda_{d\tau}  \\
e^{i(\beta_{d}+\alpha_{l})}\,\lambda_{se}  & e^{i(\beta_{d}+\beta_{l})}\, \lambda_{s\mu}  & e^{i(\beta_{d}+\gamma_{l})}\, \lambda_{s\tau}     \\
e^{i(\gamma_{d}+\alpha_{l})}\,\lambda_{be}  & e^{i(\gamma_{d}+\beta_{l})}\,\lambda_{b\mu}   &  e^{i(\gamma_{d}+\gamma_{l})}\,\lambda_{b\tau}  
\end{array}
\right)
  \overset{!}{=}
  \left(
\begin{array}{ccc}
\lambda_{de} & \lambda_{d\mu} & \lambda_{d\tau}  \\
\lambda_{se}  & \lambda_{s\mu}  & \lambda_{s\tau}     \\
\lambda_{be}  & \lambda_{b\mu}   & \lambda_{b\tau}  
\end{array}
\right)  \, , 
\end{equation}
which is clearly an over-constrained relationship; we must make assumptions about the structure of $\lambda_{dl}$ and/or the residual symmetries themselves in order to satisfy it.
We note that the simplest possibility, where the leptoquark Yukawa coupling is itself proportional to the identity matrix, does not source LNU.
In Sections \ref{sec:LQconstraints}-\ref{sec:LQexpconstraints} we discuss generic symmetry and experimental constraints on $\lambda_{dl}$, respectively, and in Section \ref{sec:otherstates} we analyze the further consequences implied by the addition of the other charged leptoquark states of \eqref{eq:LLyukFLAV}.

\subsection{Symmetry Constraints on $\lambda_{dl}$}
\label{sec:LQconstraints}
At this stage it is interesting to consider that in general the fermion masses and the leptoquark Yukawas are not simultaneously diagonal, and therefore the residual symmetry generators can at most be diagonal in the fermion mass basis or in the basis of a diagonal leptoquark Yukawa, but not in both.
The latter option we can readily exclude, as the residual symmetries would force the fermion masses to be degenerate. This is easier to see by changing the would-be diagonal residual generators into the fermion mass basis, where they are no longer diagonal, and enforcing the symmetry
\begin{equation}
\label{eq:dl_degeneracy}
X^{\dagger}_{d,l} m_{d,l} m_{d,l}^\dagger \,X_{d,l} \overset{!}{=} m_{d,l} m_{d,l}^\dagger
\end{equation}
with $X_{d,l}$ general (i.e. not the identity matrix, in which case there is no residual symmetry acting).  This forces $m_{d,l} m_{d,l}^\dagger$ (and therefore also $m_{d,l}$) to be proportional to the unit matrix.
The remaining option is to consider that the residual generators are diagonal in the mass basis as we have sketched above, and thus \eqref{eq:dl_degeneracy} holds for non-degenerate masses, with arbitrary phases in the residual symmetries as usual. The consequences are that the leptoquark Yukawas are extremely constrained, as seen explicitly in \eqref{eq:LQoverconstrain}.

Interesting solutions to \eqref{eq:LQoverconstrain} that are lepton non-universal are few in number.
From here on, we assume that the residual generators are not proportional to the identity matrix. Under this assumption, our residual symmetry distinguishes at least two generations of fermions per sector, and can therefore be considered a proper \emph{flavour} symmetry.\footnote{As discussed in Section \ref{sec:reduce}, the possibility of a residual generator $T$ being proportional to the identity matrix is viable and leads to additional patterns not considered here.} Furthermore, phenomenologically relevant patterns that can account for the $b \to s \mu \mu$ anomalies can arise only if some of the phases of $T_d$ are related. If the residual symmetry is to allow entries of the leptoquark Yukawa simultaneously in the $s$ and $b$ rows of a given column, we require in particular $\beta_d = \gamma_d$. We now further elaborate on this restriction:

\begin{enumerate}
\item If additionally either $- \alpha_l = \beta_d = \gamma_d$, $-\beta_l = \beta_d = \gamma_d$, or $- \gamma_l = \beta_d = \gamma_d$ one obtains `isolation patterns,' respectively given by
\begin{equation}
\label{eq:yukeisolation}
\lambda^{[e]}_{dl} = 
\left(
\begin{array}{ccc}
{\color{red}\lambda_{de}} & 0 & 0  \\
\lambda_{se}  & 0 & 0    \\
\lambda_{be}  & 0  & 0 
\end{array}
\right), \,\,\,\,\,
\lambda^{[\mu]}_{dl} = 
\left(
\begin{array}{ccc}
0 & {\color{red}\lambda_{d\mu}} & 0  \\
0 & \lambda_{s\mu} & 0    \\
0 & \lambda_{b\mu} & 0 
\end{array}
\right), \,\,\,\,\,
\lambda^{[\tau]}_{dl} = 
\left(
\begin{array}{ccc}
0 & 0 & {\color{red}\lambda_{d\tau}}  \\
0  & 0 & \lambda_{s \tau}     \\
0  & 0  & \lambda_{b \tau}  
\end{array}
\right)   \, . 
\end{equation}
Intriguingly, the first two of these patterns have been explored for flavoured leptoquark models before \cite{Hiller:2014yaa, Varzielas:2015iva}, due to their simplicity and phenomenological relevance. They were obtained in \cite{Varzielas:2015iva} from specific flavour symmetry models. Here we have derived them in a model independent way, simply as a consequence of the rather restrictive residual flavour symmetry. 

Furthermore, $\lambda_{d, l = e, \mu, \tau}=0$ (hence the red coloring) if $\alpha_d \neq \beta_d = \gamma_d$, according to our assumption that the residual symmetry is not proportional to the identity matrix.  It is interesting to note that, in the framework where the quark sector has a non-Abelian parent symmetry $\mathcal{G}_Q$ that breaks to $\mathcal{G}_d$ (and to $\mathcal{G}_u$), Cabibbo mixing can only be predicted by the residual symmetries if $\alpha_d \neq \beta_d = \gamma_d$ \cite{Varzielas:2016zuo}.  In other words, obtaining Cabibbo mixing in these frameworks restricts the allowed leptoquark Yukawa in precisely the same way that allows $\mathcal{R}_{K^{(\star)}}$ to also be explained by the residual symmetry!\footnote{\label{footnote}Given that we are presenting model independent results, we note that this connection between the Cabibbo angle and $\mathcal{R}_{K^{(\star)}}$ may be lost in frameworks where the $\mathcal{G}_d$ does not arise from $\mathcal{G}_Q$ in this manner.}

\item If the lepton phases are also related to one another, one can allow entries in more than one column of the leptoquark Yukawa, and the leptoquark coupling must have, for a given quark row, at least one zero. This is also consistent with non-trivial leptonic mixing being predicted by the residual symmetry in the framework where the lepton sector has a non-Abelian parent symmetry $\mathcal{G}_\mathcal{L}$ that breaks to $\mathcal{G}_e$ (and to $\mathcal{G}_\nu$). Note also that at least two non-zero entries in the same row of the leptoquark coupling are required for LFV processes such as $\mu \to e \gamma$ --- in this case some $\lambda_{qe}$ and $\lambda_{q\mu}$ (same $q$) are needed.  For example, taking $\alpha_{l} = \beta_{l} = -\beta_{d} =-\gamma_{d} $, $\alpha_{l} = \gamma_{l} = -\beta_{d} =-\gamma_{d}$ or $\beta_{l} = \gamma_{l} = -\beta_{d} =-\gamma_{d}$ one respectively finds

\begin{equation}
\label{eq:yukem}
\lambda^{[e\mu]}_{dl} = 
\left(
\begin{array}{ccc}
0  & 0 & 0  \\
\lambda_{se} & \lambda_{s\mu} & 0    \\
 \lambda_{be} & \lambda_{b\mu} & 0 
\end{array}
\right), \,\,\,\,\,\,
\lambda^{[e\tau]}_{dl} = 
\left(
\begin{array}{ccc}
0  & 0 & 0 \\
\lambda_{se} & 0 & \lambda_{s\tau}    \\
 \lambda_{be} & 0 & \lambda_{b\tau} 
\end{array}
\right), \,\,\,\,\,\,
\lambda^{[\mu\tau]}_{dl} = 
\left(
\begin{array}{ccc}
0 & 0  & 0  \\
0 &  \lambda_{s\mu}  & \lambda_{s\tau}    \\
0 & \lambda_{b\mu} & \lambda_{b\tau} 
\end{array}
\right)  \, , 
\end{equation}
where we have again insisted that $T_{d}$ not be proportional to the identity.  We note also that the $\lambda_{dl}^{[\mu\tau]}$ pattern has a vanishing first row and first column, making it analogous to one of the cases considered in \cite{Varzielas:2015iva}, which originated from an $SU(3)_F$ family symmetry (namely \cite{deMedeirosVarzielas:2005ax} or similar constructions accounting for the reactor angle, e.g. \cite{deMedeirosVarzielas:2017sdv}).

\item Finally, although we do not permit $\alpha_{d} \neq \beta_{d} = \gamma_{d}$, we still have freedom to allow one additional entry in the empty columns of \eqref{eq:yukem} by setting $\alpha_{d} =-\lbrace \alpha_{l}, \beta_{l}, \gamma_{l}  \rbrace$, with the three different solutions respectively corresponding to an augmented first, second, and third column:
\begin{equation}
\label{eq:yukemt}
\lambda^{[e\mu1]}_{dl} = 
\left(
\begin{array}{ccc}
0  & 0 & \lambda_{d\tau}  \\
\lambda_{se} & \lambda_{s\mu} & 0    \\
 \lambda_{be} & \lambda_{b\mu} & 0 
\end{array}
\right), \,\,\,\,\,\,
\lambda^{[e1\tau]}_{dl} = 
\left(
\begin{array}{ccc}
0  & \lambda_{d\mu}  & 0  \\
\lambda_{se} & 0 & \lambda_{s\tau}    \\
 \lambda_{be} & 0 & \lambda_{b\tau} 
\end{array}
\right), \,\,\,\,\,\,
\lambda^{[1\mu\tau]}_{dl} = 
\left(
\begin{array}{ccc}
\lambda_{de}  & 0  & 0 \\
0 &  \lambda_{s\mu}  & \lambda_{s\tau}    \\
0 & \lambda_{b\mu} & \lambda_{b\tau} 
\end{array}
\right) \, . 
\end{equation}
\end{enumerate}

In summary, under the simple assumptions that \emph{1)} residual flavour symmetries in the charged lepton and down quark sectors leave the extended Yukawa sector invariant, and \emph{2)} that the new scalar leptoquark explains observed B-decay anomalies, we force the possible $d-l$ leptoquark coupling into patterns with specific column structures given by \eqref{eq:yukeisolation}-\eqref{eq:yukemt}.

\subsection{Experimental Constraints on $\lambda_{dl}$}
\label{sec:LQexpconstraints}
A wide variety of lepton flavour violating (LFV) and LNU tests can be employed to constrain the matrix elements of $\lambda_{dl}$.  In Table \ref{tab:LFV} we give the list of relevant LFV bounds, including projected future sensitivities.  Of course we can also recast the hints of $\mathcal{R}_{K^{(*)}}$ into our framework, which if explained by a $\Delta_3$, $\Delta_3^\mu$, or $\Delta_1^\mu$ leptoquark,  gives the following constraint \cite{Varzielas:2015iva, Hiller:2017bzc, Hiller:2018wbv}:
\begin{equation}
\label{eq:RKstarexp}
\lambda_{b\mu} \lambda_{s\mu}^* - \lambda_{be}\lambda_{se}^* \simeq \frac{1.1 M^2}{(35~\text{TeV})^2} \, .
\end{equation}
Furthermore, a strong upper bound on the same couplings is obtained from the $B_s - \bar{B}_s$ mixing phase, which can be expressed as \cite{Varzielas:2015iva}
\begin{equation}
\label{eq:BBar}
(\lambda_{se}\lambda_{be}^*+\lambda_{s\mu}\lambda_{b\mu}^*+\lambda_{s\tau}\lambda_{b \tau}^*)^2 \lesssim \frac{M^2}{(17.3 \, \text{TeV})^2} \, .
\end{equation}

\begin{table} [t]
\centering
\small
{\renewcommand{\arraystretch}{1.5}
\begin{tabular}{|c||c|c|c|}
\hline
 Observable & Current 90 \% CL Limit  & Constraint & Future Sensitivity \\
\hline
\hline
${\cal{B}}(\mu \to e \gamma) $ & $ 5.7 \cdot 10^{-13}$  \cite{Adam:2013mnn} & $|  \lambda_{qe}\lambda_{q\mu}^* | \lesssim   \frac{M^2}{(34 {\rm TeV})^2} $ & $6 \cdot 10^{-14}$ \cite{Baldini:2013ke}\\
${\cal{B}}(\tau \to e \gamma)$ & $1.2 \cdot 10^{-7}$ \cite{Hayasaka:2007vc} & $|  \lambda_{qe}\lambda_{q\tau}^* |  \lesssim   \frac{M^2}{(0.6 {\rm TeV})^2} $ &\\
${\cal{B}}(\tau \to \mu \gamma)$ & $4.4 \cdot 10^{-8}$   \cite{Aubert:2009ag}  &  $|  \lambda_{q\mu}\lambda_{q\tau}^* |  \lesssim   \frac{M^2}{(0.7 \, {\rm TeV})^2}$ & $5 \cdot 10^{-9}$ \cite{Aushev:2010bq}\\
${\cal{B}}(\tau \to \mu \eta)$ & $6.5 \cdot 10^{-8}$   \cite{Miyazaki:2007jp} & $|  \lambda_{s \mu}\lambda_{s \tau}^* | \lesssim   \frac{M^2}{(3.7 \,  {\rm TeV})^2}$ & $2 \cdot 10^{-9}$ \cite{Aushev:2010bq}\\
${\cal{B}}(B \to K  \mu^\pm e^\mp)$ &  $ 3.8 \cdot 10^{-8}$ \cite{Aubert:2006vb} & $\sqrt{ |  \lambda_{s \mu }\lambda_{b e}^* |^2+ |  \lambda_{b \mu }\lambda_{se}^* |^2} \lesssim   \frac{M^2}{(19.4\,  {\rm TeV})^2} $ & \\
${\cal{B}}(B \to K  \tau^\pm e^\mp) $ & $3.0 \cdot 10^{-5}$ \cite{Agashe:2014kda} & $\sqrt{ |  \lambda_{s \tau }\lambda_{b e}^* |^2+ |  \lambda_{b \tau }\lambda_{se}^* |^2} \lesssim   \frac{M^2}{(3.3 \,  {\rm TeV})^2} $ & \\
${\cal{B}}(B \to K  \mu^\pm \tau^\mp) $ & $ 4.8 \cdot 10^{-5}$  \cite{Agashe:2014kda} & $\sqrt{ |  \lambda_{s \mu }\lambda_{b \tau}^* |^2+ |  \lambda_{b \mu }\lambda_{s\tau}^* |^2} \lesssim   \frac{M^2}{(2.9 \,  {\rm TeV})^2}$ &  \\
${\cal{B}}(B \to \pi  \mu^\pm e^\mp)$ &  $ 9.2 \cdot 10^{-8}$ \cite{Aubert:2007mm} & $\sqrt{ |  \lambda_{d \mu }\lambda_{b e}^* |^2+ |  \lambda_{b \mu }\lambda_{de}^* |^2} \lesssim   \frac{M^2}{(15.6\,  {\rm TeV})^2} $ & \\
\hline
\end{tabular}}
\caption{Bounds on the leptoquark couplings from LFV processes ($q=d,s,b$). Belle II projections  \cite{Aushev:2010bq} are for  $50 \, ab^{-1}$. 
For ${\cal{B}}(\tau \to  \mu \eta)$ we ignored possible cancellations with $ \lambda_{d \mu}\lambda_{d \tau}^*$, see {\it e.g.,} \cite{Dorsner:2011ai}.
As in \cite{Davidson:1993qk}, we ignored tuning between leading order diagrams in the amplitudes of $\ell \to \ell^\prime  \gamma$.}
\label{tab:LFV}
\end{table}

In order to succinctly summarize the combined implications of Table \ref{tab:LFV} and \eqref{eq:RKstarexp}-\eqref{eq:BBar}, we follow \cite{Hiller:2018wbv} and utilize a special parameterization for the coupling that captures the interesting splittings between lepton and quark species.  In particular, we employ
\begin{equation}
\label{eq:LQparagen}
\lambda_{dl} \overset{!}{\equiv}
\lambda_{0}
\left(
\begin{array}{ccc}
\rho_{d}\,\kappa_{e} & \rho_{d} \, \kappa_{\mu} & \rho_{d}\, \kappa_{\tau}  \\
\rho \, \kappa_{e}   &\rho \, \kappa_{\mu} & \rho \, \kappa_{\tau}     \\
\kappa_{e}   & \kappa_{\mu}   &  \kappa_{\tau} 
\end{array}
\right)  ,
\end{equation}  
where $\lambda_0$ is an overall scale-setting parameter, $\rho$ and $\rho_d$ encode splittings between quark species, and $\kappa_{l}$ similarly encode lepton splittings.   The defining values and/or implied experimental bounds for the parameters are given for each relevant, symmetry-allowed $\lambda_{dl}$ pattern in Table \ref{tab:exp2}.\footnote{It is clear that the three-columned matrices of \eqref{eq:yukemt} will not fit into \eqref{eq:LQparagen}.  However, as is demonstrated in Section \ref{sec:threecolumn}, these matrices reduce to the two-columned patterns of \eqref{eq:yukem} upon considering SU(2) rotations.  We therefore do not need to consider them here.}  

A few comments are in order regarding \eqref{eq:LQparagen} .  First of all, its imposition represents a trivial rewrite of the original isolation patterns, as the absence of a lepton splitting parameter $\kappa_{l}$ implies that $\lambda_0 \equiv \lambda_{bl}$ and $\rho \equiv \lambda_{sl}/\lambda_{bl}$ without loss of generality.  Then the combined constraints on $\mathcal{R}_{K^{(\star)}}$ and $B_s$-$\bar{B}_s$ mixing imply an upper bound of $\mathcal{O}(50)$ TeV for the leptoquark mass $M$ and a very weak bound on the quark splitting parameter $\rho$ as seen in Table \ref{tab:exp2} \cite{Hiller:2014yaa}. Note that as the isolation patterns are not LFV, Table \ref{tab:LFV} gives no additional information.
On the other hand, \eqref{eq:LQparagen} does imbed certain biases into our parameterization of the relevant data for two-columned patterns.  After all, \eqref{eq:LQparagen} reduces the four complex parametric degrees of freedom in $\lambda_{dl}^{[l_{1}l_{2}]}$ to three, and so it cannot be entirely generic.  Indeed, its form implies that lepton splittings are universal for any given quark species (divide across columns) and that, similarly, quark splittings are universal for any given lepton species (divide across rows).  Furthermore, as we will see below, \eqref{eq:LQparagen} as quantized in Table \ref{tab:exp2} also implies that leptoquark couplings mimic SM flavour hierarchies, with couplings to heavier fermions larger than those of lighter ones.  We refer to this parameterization as the `normal hierarchy' scenario, as its assumptions are motivated by our desire to utilize flavour symmetries to structure \emph{both} SM and leptoquark Yukawa sectors.  Regardless, we caution that some derivations in Section \ref{sec:otherstates} are sensitive to this choice, and so a treatment of inverted hierarchies, while beyond the scope of this introductory paper on the simplified model space, will be pursued in a more exhaustive phenomenological survey to appear in a future publication. 

Proceeding, the bounds from LFV processes $\mathcal{B} \left(l_{1} \rightarrow l_{2} \gamma \right)$ in Table \ref{tab:LFV} provide a stronger upper limit on $\rho$ for two-columned matrices.  To see this simply expand the constraint in the parameterization for multiple quark rows:
\begin{align}
\label{eq:rho1}
\vert \lambda_{sl_{1}} \lambda_{sl_{2}}^{\star} \vert \equiv \vert \lambda_{0} \lambda_{0}^{\star} \kappa_{l} \rho \rho^{\star} \vert =  \vert \lambda_{0}^{2} \kappa_{l} \rho^{2} \vert &\lesssim \frac{M^2}{(x\,TeV)^2} \, , \\
\label{eq:rho2}
\vert \lambda_{bl_{1}} \lambda_{bl_{2}}^{\star} \vert \equiv \vert \lambda_{0}^{2} \kappa_{l} \vert &\lesssim \frac{M^2}{(x\,TeV)^2} \, ,
\end{align}
such that, if both \eqref{eq:rho1} and \eqref{eq:rho2} are true, $\vert \rho \vert ^{2} \le 1$.  This conclusion holds regardless of $x$, the experimental bound on the leptoquark mass $M$, and regardless of whether normal or inverted hierarchies (corresponding to different row/column placements for $\rho$, $\kappa_{l}$) are assumed.  Similarly, a powerful upper bound on $\kappa_{l}$ can be obtained from $B_s$-$\bar{B}_s$ mixing, which implies
\begin{equation}
\label{eq:BBbound1}
\rho^{2}\cdot \left( \vert \lambda_0 \vert^{2} \vert \kappa_{l} \vert^2+  \vert \lambda_0 \vert^{2} \right)^{2} \lesssim \frac{M^{2}}{\left(17.3 \,TeV \right)^{2}} \, .
\end{equation}
We now observe that the bracketed term on the LHS is a quadratic polynomial of the form $(a+ b)^{2} = a^2 + b^2 + 2 a b$
where, importantly, the three terms on the RHS of this expression are by definition positive quantities.  Hence, with $\rho^{2} \ge 0$, 
\eqref{eq:BBbound1} demands that any of the positive definite quantities on the LHS are \emph{themselves} less than the mixing bound,
\begin{equation}
\left(  \vert \lambda_0 \vert^{2} \vert \kappa_{l} \vert^2 \right)^2 \lesssim \frac{M^{2}}{\left(17.3 \,TeV \right)^{2}}, \,\,\,\,\,\,\,
\left(  \vert \lambda_0 \vert^{2}\right)^2 \lesssim \frac{M^{2}}{\left(17.3 \,TeV \right)^{2}}, \,\,\,\,\,\,\,\text{and} \,\,\,\,\,\,\,
2\left( \vert \lambda_0 \vert^{2} \right)^2 \vert \kappa_{l} \vert^2 \lesssim \frac{M^{2}}{\left(17.3 \,TeV \right)^{2}} \, ,
\end{equation}
and therefore, using the latter two terms, we immediately derive that
\begin{equation}
\vert \kappa_{i} \vert \le1/\sqrt{2} \simeq 0.71 \, .
\label{eq:geobound}
\end{equation}
This is again independent of the actual leptoquark mass bound and the imposition of normal or inverted hierarchies (so long as $\rho^{2} \ge 0$ holds).  Note that a stronger bound of $\kappa_{e} \lesssim 1/2$ was given in prior analyses considering  patterns with an electron splitting parameter \cite{Varzielas:2015iva}.  This can be obtained via LFV constraints, but requires further assumptions.  Hence we use the conservative geometric bound of $1/\sqrt{2}$ in what follows, which in any event does not affect our conclusions in Section \ref{sec:otherstates}, thanks to current sensitivities  in CKM and PMNS mixing matrix elements.

\begin{table} [t]
\centering
{\renewcommand{\arraystretch}{2.0}
\begin{tabular}{|c||c|c|c|c|c|}
\hline
Pattern & $\rho_{d}$ & $\rho$ & $\kappa_{e}$ & $\kappa_{\mu}$ & $\kappa_{\tau}$ \\
\hline
\hline
e-isolation & 0  & $ 10^{-4} \lesssim \rho \lesssim 10^{4} $ & 1 & 0 & 0 \\
\hline
$\mu$-isolation & 0 & $ 10^{-4} \lesssim \rho \lesssim 10^{4} $ & 0 & 1 & 0 \\
\hline
\hline
e-$\mu$ & 0 & $ 10^{-4} \lesssim \rho \lesssim 1$  & $ \kappa_{e} \le 1/\sqrt{2}$  & 1 & 0 \\
\hline
e-$\tau$  & 0 & $ 10^{-4} \lesssim \rho \lesssim 1$ & $ \kappa_{e} \le1/\sqrt{2}$ & 0 & 1\\
\hline
$\mu$-$\tau$ & 0 & $ 10^{-4} \lesssim \rho \lesssim 1$ & 0 & $\kappa_\mu \le 1/\sqrt{2}$ & 1 \\
\hline
\end{tabular}}
\caption{Defining values and/or experimentally bound ranges for the parameterization of \eqref{eq:LQparagen}, in the normal hierarchy scenario.  As noted in the text, three-columned patterns are reduced to their two-columned cousins upon SU(2) symmetry considerations (cf. Section \ref{sec:threecolumn}), and hence we do not need to fit such matrices. }
\label{tab:exp2}
\end{table}

\subsection{Collider Implications}
\label{sec:collider}

The phenomenology of leptoquarks at proton-proton colliders has been explored in multiple previous works (see e.g.  \cite{Dorsner:2016wpm,Diaz:2017lit, Dorsner:2017ufx, Dorsner:2018ynv, Hiller:2018wbv, Schmaltz:2018nls}).  While it is beyond our current scope to make detailed predictions for collider observables, we will briefly review the qualitative conclusions of those papers here and discuss how our models are related to those previously studied.

The collider channels most sensitive to leptoquarks are those where a generic vector or scalar leptoquark $\Delta$ is $1)$ pair produced and decays to two quarks and two leptons $(pp \rightarrow \Delta \bar{\Delta} \rightarrow (LQ)(LQ))$, $2)$ singly produced and decays to a di-lepton + quark (jet) final state $(pp \rightarrow L^{+}L^{-} j)$ or $3)$ mediates a t-channel exchange between quarks and leptons, leading to a Drell-Yan (DY)-like final state $(pp \rightarrow L^{+}L^{-})$.  In the limit of small coupling $y$, pair production dominates the cross section which is $\propto g_{s}^2$, the strong coupling, and LHC cross section constraints only lead to bounds on the leptoquark mass $M$ (at least for scalars --- see the caveat below).  When the coupling is allowed to be larger, however, contributions from single production and DY-like diagrams can become relevant/complementary, allowing constraints to be given on the $y-M$ plane.  The most detailed analyses of this type to date have been done for the so-called `Minimal Leptoquark' Models of \cite{Diaz:2017lit,Schmaltz:2018nls}, where $\Delta$ only couples to one combination of quark and lepton species, and where the leptoquark decay widths are approximated by
\begin{equation}
\label{eq:decaywidths}
\Gamma_{S} \left(\Delta_{S} \rightarrow Q L\right) \approx \frac{y^2}{16 \pi} M_{S}, \,\,\,\,\,\,\,\,\,\,\Gamma_{V} \left(\Delta_{V} \rightarrow Q L\right) \approx \frac{y^2}{24 \pi} M_{V},
\end{equation} 
where $S$ and $V$ respectively denote `scalar' and `vector'.  The take-away `rule-of-thumb' from these studies (also see \cite{Dorsner:2018ynv, Hiller:2018wbv}) is that pair production constrains the scalar leptoquark mass $M_S$ with almost no dependence on the couplings to fermions while DY-like searches limit the size of the couplings $y$.  For example, considering $\Delta_3$ and using the bounds from \cite{Diaz:2017lit}, we estimate $M_S > 1.4~\text{TeV}$ when $\Delta_3$ couples mostly to $b \mu$ and $M_S > 1.0~\text{TeV}$ when $\Delta_3$ couples mostly to $b \tau$. Bounds from single production exclude masses $M_S > 1.5~\text{TeV}$ for larger coupling strengths.  Vector leptoquark mass bounds are typically larger but also more model dependent, due to the need to UV-complete such models \cite{Schmaltz:2018nls}.

Of course in our case $\Delta$ couples to more than one flavour of quark and lepton, as can already be seen in \eqref{eq:yukeisolation}-\eqref{eq:yukemt} but more explicitly in Appendix \ref{app:final}, which incorporates the conclusions from Sections \ref{sec:otherstates}-\ref{sec:otherstatesV1V3}. Yet ours is also a flavour-symmetric approach, and as expected we typically find that $\Delta_3$ couples dominantly to a specific flavour of quark and/or lepton, with the remaining couplings suppressed by e.g. CKM elements.  In this case $y^2 \approx c \lambda_{QL}^2$ in \eqref{eq:decaywidths} with $\lambda_{QL}$ the largest coupling and $c$ a coefficient depending on the hypercharge of $\Delta_3$ \cite{Hiller:2018wbv}.  That is, the total decay width is approximated by that coming only from the dominant coupling, and therefore one expects the rough magnitude of $M_S$ reported above to also hold in our simplified models.  Indeed, the authors of \cite{Hiller:2018wbv} explored the collider phenomenology of multiple models with flavour-symmetry inspired $\lambda_{QL}$ that incorporated SM-like flavour suppressions, including the generic pattern in \eqref{eq:LQparagen}, with $\kappa_{\mu} = 1$ and $\kappa_\tau \sim 1$.  They found that, in the narrow width approximation, the parametric signal strength of $pp \rightarrow (b/j) l l^{\prime}$ sourced from single $\Delta_3^{-4/3}$ production was (e.g.) $\lambda_0^2/2$, $\lambda_0^2 \kappa_e^2/2$, $\lambda_0^2 \rho^2/2$, and $\lambda_0^2 \rho^2 \kappa_e^2/2$ for $b\mu\mu$, $b e \mu$, $j \mu \mu$, and $\j e \mu$ final states, respectively.  They also concluded that if $\rho$, $\kappa_e \ll 1$ when $\lambda_0$ is substantial, then $M_S$ can be $\mathcal{O}\left(TeV\right)$.  As noted above, it would be interesting to explore these bounds and explicit cross section predictions given the matrices in Appendix \ref{app:final}.  After all, we do not generically observe large flavour suppressions across rows (but we do across columns).  The important point is that our simplified models still imitate other flavour symmetry approaches, and so we expect that a detailed collider study would simply give detailed (but sensible) bounds on $\lambda_{bl}/M$ (cf. Appendix \ref{app:final}).
\section{$SU(2)$ analysis for the scalar $\Delta_{3}$}
\label{sec:otherstates}

Upon $SU(2)$ decomposition, all  couplings in \eqref{eq:LLyukFLAV}-\eqref{eq:LLyukFLAV2} coming from leptoquarks with different electric charge must be considered.  We will first address this for the scalar triplet, and then see that the discussion easily generalizes to the vector states in Sec. \ref{sec:otherstatesV1V3}.

For $\Delta_{3}$, the full Yukawa sector Lagrangian, in the mass basis of the SM fermions, then reads
\begin{align}
\nonumber
\mathcal{L} \,\,\, \supset \,\,\,&\frac{1}{2} \bar{\nu}^{c}_{L}\, m_{\nu}\, \nu_{L} +  \bar{l}_{L}\, m_{l}\,E_{R}   +  \bar{d}_{L}\, m_{d} \, d_{R}   + \bar{u}_{L} \, m_{u} \, u_{R}   \\
\nonumber
&+\, \bar{d}^{C}_{L} \,\lambda_{dl}\, l_{L} \, \Delta^{4/3}_{3} + \bar{d}^{C}_{L} \,\lambda_{d\nu}\, \nu_{L} \, \Delta^{1/3}_{3} + \bar{u}^{C}_{L} \,\lambda_{ul}\, l_{L} \, \Delta^{1/3}_{3} + \bar{u}^{C}_{L} \,\lambda_{u\nu}\, \nu_{L} \, \Delta^{-2/3}_{3} \\
\nonumber
&+\, \bar{d}^{C}_{L} \,\lambda_{du}\, u_{L} \, \Delta^{1/3,\star}_{3} + \bar{d}^{C}_{L} \,\lambda_{dd}\, d_{L} \, \Delta^{-2/3,\star}_{3} + \bar{u}^{C}_{L} \,\lambda_{uu}\, u_{L} \, \Delta^{4/3,\star}_{3} + \bar{u}^{C}_{L} \,\lambda_{ud}\, d_{L} \, \Delta^{1/3,\star}_{3} \\
\label{eq:LQYUKfull}
&+\, \text{h.c.}
\end{align}
where $m_{a}$, with $a \in \lbrace u,d,l, \nu \rbrace$, are all diagonal matrices of mass eigenvalues, and the $\lambda_{(QL,QQ)}$ matrices are analogous to \eqref{eq:genericyuk}.  Because we are in the mass basis of all of the fermions,  an analogous equation \eqref{eq:LQoverconstrain} arises for each leptoquark Yukawa term if there are residual symmetries in each SM fermion sector,\footnote{In this paper we only consider the case where a single generator $T_{\nu}$ is active in the neutrino sector, with (a priori) three independent phases.}
\begin{equation}
\label{eq:LQoverconstrainGEN}
T^{T}_{Q}\, \lambda_{QL}\,T_{L} \overset{!}{=} \lambda_{QL} \,\,\,\, \forall \,\,\,\, \lbrace Q, L \rbrace ,
\end{equation}
where the $T$ matrices are again diagonal residual symmetry generators with three phases, and with an equivalent equality holding for $\lambda_{QQ}$ couplings.  Hence, we must make the same types of considerations discussed in Section \ref{sec:LQconstraints} for \emph{each} coupling matrix, which we discuss pattern by pattern below.  We also still demand that the residual symmetries act non-trivially in each sector, i.e. that $T_{Q,L}$ are not proportional to the identity matrix --- cf. Section \ref{sec:reduce} for a relaxation of this assumption.

However, we are further constrained by the fact that the eight $\lambda_{(QL,QQ)}$ are not independent, as they are sourced from the two original couplings, $y^{LL}_{3}$ and $z_{3}^{LL}$. Terms originating from the quark-lepton operator can therefore all be normalized to $\lambda_{dl}$, the term for which we have some phenomenological insight given the anomalies in the $B$-decay data, and those from the diquark operator can be normalized to $\lambda_{du}$.  We find, using that 
\begin{equation}
\label{eq:CKMandPMNS}
U_{CKM} \equiv U^{\dagger}_{u} \, U_{d}, \,\,\,\,\,\,\,\,\,\,U_{PMNS} \equiv U^{\dagger}_{l} \, U_{\nu},
\end{equation}
the following relationships between the different charged leptoquark couplings:
\begin{align}
\nonumber
\lambda_{d \nu} &= \frac{1}{\sqrt{2}} \lambda_{dl}\,U_{PMNS}, \,\,\,\,\,\, &&\lambda_{ul} = \frac{1}{\sqrt{2}} U^{\star}_{CKM}\,\lambda_{dl}, \,\,\,\, &&\lambda_{u \nu} = - U^{\star}_{CKM}\,\lambda_{dl}\,U_{PMNS}, \\
\label{eq:LQrelations}
\lambda_{d d} &= \sqrt{2}\, \lambda_{du}\,U_{CKM}, \,\,\,\,\,\, &&\lambda_{uu} = -\sqrt{2}\, U^{\star}_{CKM}\,\lambda_{du}, \,\,\,\, &&\lambda_{ud} = U^{\star}_{CKM}\,\lambda_{du} \, U_{CKM}. 
\end{align}
We will see below that \eqref{eq:LQrelations} has severe implications given the viable forms of $\lambda_{dl}$.  We will also find that, upon considering global fits on the CKM \cite{Tanabashi:2018oca} and PMNS \cite{Esteban:2016qun,NuFit} matrix elements given by
\begin{equation}
\label{eq:num}
|U_{CKM}| \simeq
\renewcommand{\arraystretch}{1.4}
\left(
\begin{array}{ccc}
  \left(0.97456 \atop 0.97436 \right)  &  \left(0.22496 \atop 0.22408 \right)  & \left(0.00377 \atop 0.00353 \right)  \\
  \left(0.22482 \atop 0.22394 \right) &  \left(0.97369 \atop 0.97348 \right)  &  \left(0.04290 \atop 0.04138 \right) \\
  \left(0.00920 \atop 0.00873 \right) &  \left(0.04207 \atop 0.04059 \right)  &  \left(0.999137 \atop 0.999073 \right) \\
\end{array}
\right), \,\,\,\,\,\,\,\,
|U_{PMNS}| \simeq
\renewcommand{\arraystretch}{1.4}
\left(
\begin{array}{ccc}
  \left(0.844 \atop 0.799 \right)  &  \left(0.582 \atop 0.516 \right)  & \left(0.156 \atop 0.141 \right)  \\
  \left(0.494 \atop 0.242 \right) &  \left(0.678 \atop 0.467 \right)  &  \left(0.774 \atop 0.639 \right) \\
  \left(0.521 \atop 0.284 \right) &  \left(0.695 \atop 0.490 \right)  &  \left(0.754 \atop 0.615 \right) \\
\end{array}
\right)\,,
\end{equation}
that further constraints on acceptable leptoquark patterns arise.
Henceforth we use the shorthand notation
\begin{equation}
U^{ij}_{PMNS} = U_{ij},\,\,\,\,\,\,\,\,\,\,
(U^{ij}_{CKM})^{\star} = V_{ij},
\end{equation}
where for $U$ we use $i,j=1,2,3$ and for $V$ we use as indices $i=u,c,t$ and $j=d,s,b$.

In what follows we analyze the implications of \eqref{eq:LQparagen},  \eqref{eq:LQoverconstrainGEN}, \eqref{eq:LQrelations} and \eqref{eq:num} on the few $\lambda_{dl}$ that account for $\mathcal{R}_{K^{(\star)}}$ and are allowed by the residual symmetries. From the purely flavour symmetric perspective, in the quark-lepton sector we find as before that the solutions for each equation implied by \eqref{eq:LQoverconstrainGEN} give matrices analogous to those in Section \ref{sec:LQconstraints}, up to permutations of rows and columns.  So long as $T_{Q,L}$ are symmetries of the Lagrangian, this statement holds \emph{regardless} of the relationships implied by \eqref{eq:LQrelations} --- it is true simply by virtue of the phase constraints in $T_{Q}$ and $T_{L}$.

We now treat each acceptable pattern of $\lambda_{dl}$ case by case by deriving the combined symmetry and experimental constraints, and showing the resultant matrices explicitly.  We then consider the implications of these constraints on the diquark operators, and further discuss whether additional restrictions must be imposed to forbid proton decay.

\subsection{Isolation Patterns:  $\lambda_{dl} = \lambda^{[e,\mu,\tau]}_{dl}$}
\label{sec:isolation}

We first treat the case where $\lambda_{dl}$ is in an isolation pattern.  First considering electron isolation, the explicit matrix for $\lambda_{d\nu}$ is given by
\begin{equation}
\label{eq:dnu}
\lambda_{d\nu}^{[e]} =
\frac{1}{\sqrt{2}} \left(
\begin{array}{ccc}
0 & 0 & 0 \\
U_{11} \lambda_{se} & U_{12} \lambda_{se}  & U_{13}\lambda_{se}  \\
U_{11} \lambda_{be} & U_{12}\lambda_{be} & U_{13} \lambda_{be} 
\end{array}
\right)
\end{equation}
and we have  set $\lambda_{de} = 0$ as required. Muon or tau isolation simply implies $e \rightarrow \lbrace \mu, \tau \rbrace$ and $U_{1i} \rightarrow \lbrace U_{2i}, U_{3i} \rbrace$ in \eqref{eq:dnu}, respectively.
Explicitly, one finds
\begin{equation}
\label{eq:dnu_mu}
\lambda_{d\nu}^{[\mu]} =
\frac{1}{\sqrt{2}} \left(
\begin{array}{ccc}
0 & 0 & 0 \\
U_{21} \lambda_{s\mu} & U_{22} \lambda_{s\mu}  & U_{23}\lambda_{s\mu}  \\
U_{21} \lambda_{b\mu} & U_{22}\lambda_{b\mu} & U_{23} \lambda_{b\mu} 
\end{array}
\right), \,\,\,\,\,\,\,\,\,\,
\lambda_{d\nu}^{[\tau]} =
\frac{1}{\sqrt{2}} \left(
\begin{array}{ccc}
0 & 0 & 0 \\
U_{31} \lambda_{s\tau} & U_{32} \lambda_{s\tau}  & U_{33}\lambda_{s\tau}  \\
U_{31} \lambda_{b\tau} & U_{32}\lambda_{b\tau} & U_{33} \lambda_{b\tau} 
\end{array}
\right)
\end{equation}
for muon and tau isolation.

We first notice that the $\lambda_{d \nu}$ coupling is not allowed to take an isolation pattern, as this would force all entries (in all couplings) to zero, since only one matrix element of $U_{PMNS}$ is measured to be small.  This then leaves us with the multi-column options, where we further read off that the $\lambda_{d\nu}^{1i}$ row is zero (a consequence of $\beta_{d} = \gamma_{d}$).  Next, we need to set two matrix elements in one column to zero in \eqref{eq:dnu}-\eqref{eq:dnu_mu}.  This demand is particularly powerful because, regardless of whether $\lambda_{dl}$ isolates electrons, muons, or tauons it requires either $\lambda_{sl} = \lambda_{bl} = 0$ \emph{or} a single matrix element of $U_{PMNS}$ to zero.  The former option sets all leptoquark Yukawa couplings to zero, so is not interesting.  Hence, our residual flavour symmetry is forcing us to a limit where $U_{PMNS}$ has a null matrix element, which is experimentally excluded.  \emph{We can therefore conclude that the combined SU(2) and flavour constraints do not permit isolation patterns for $\lambda_{dl}$ to first approximation.}

However, as a pedagogical preparation for later Sections, and because the limit $U_{13} = 0$ is still a reasonable approximation to data (and can be the starting point for flavour models \cite{Sierra:2014hea}), we continue with our analysis.  Allowing $U_{13} = 0$ (but no other null matrix elements), we then find that $\lambda_{dl} = \lambda^{[e]}_{dl}$, as all other isolation patterns (cf. (\ref{eq:dnu_mu})) would require some other mixing element to be zero. We conclude that $\lambda_{dl} = \lambda^{[e]}_{dl}$, $\lambda_{d\nu} = \lambda_{d \nu}^{[e3]}$ (with $U^{13}_{PMNS} = 0$), and $\alpha_{\nu} = \beta_{\nu} = -\beta_{d} = - \gamma_{d}$ (the superscript `3' denotes that the third column vanishes when coupling to $\nu$). 

We now consider the $\lambda_{ul}$ coupling corresponding the electron isolation $\lambda_{dl}^{[e]}$:
\begin{equation}
\label{eq:ul}
\lambda_{ul} =
\frac{1}{\sqrt{2}} \left(
\begin{array}{ccc}
V_{ub} \lambda_{be} +  V_{us} \lambda_{se} & 0 & 0 \\
V_{cb} \lambda_{be} +  V_{cs} \lambda_{se} & 0 & 0  \\
V_{tb} \lambda_{be} +  V_{ts} \lambda_{se} & 0 & 0 
\end{array}
\right) \, . 
\end{equation}
We see that $\lambda_{ul}$ is naturally of an isolation pattern form, with the following constraint on one of its matrix elements:
\begin{equation}
\label{eq:ulconstraint}
\frac{\lambda_{se}}{\lambda_{be}} = - \frac{V_{ib}}{V_{is}}  \, , 
\end{equation} 
where $i \in \lbrace u,c,t \rbrace$ and its specific value is determined by the phases of $T_{u}$. The three solutions are either
$\beta_{u} = \gamma_{u} = - \alpha_{l}$ with $\frac{\lambda_{se}}{\lambda_{be}} = - \frac{V_{ub}}{V_{us}}$ ($i=1$),
$\alpha_{u} = \gamma_{u} = - \alpha_{l}$ with $\frac{\lambda_{se}}{\lambda_{be}} = - \frac{V_{cb}}{V_{cs}}$  ($i=2$)
or
$\alpha_{u} = \beta_{u} = - \alpha_{l}$ with $\frac{\lambda_{se}}{\lambda_{be}} = - \frac{V_{tb}}{V_{ts}}$  ($i=3$).
We label these couplings, upon the application of \eqref{eq:ulconstraint}, $\lambda_{ul}^{[eA]}$, $\lambda_{ul}^{[eB]}$ and $\lambda_{ul}^{[eC]}$, respectively denoting with superscripts $A,B,C$ that the first, second, or third row vanishes. Note that each solution to \eqref{eq:ulconstraint} is also communicated back to $\lambda_{dl}$ and $\lambda_{d \nu}$, in the sense that the $\frac{\lambda_{se}}{\lambda_{be}}$ is now related to ratios of CKM elements. We respectively denote the resulting matrices as $\lambda_{dl}^{[e3i]}$ and $\lambda_{d\nu}^{[e3i]}$ (with $i=A,B,C$).

Finally, we write down $\lambda_{u\nu}$:
\begin{align}
\label{eq:unu1}
\lambda_{u\nu}^{[e3A]} &=
\lambda_{be} \left(
\begin{array}{ccc}
0 & 0  & 0    \\
U_{11} \left(\frac{V_{ub}V_{cs}}{V_{us}} - V_{cb} \right)  & U_{12} \left(\frac{V_{ub}V_{cs}}{V_{us}} - V_{cb} \right)   & 0      \\
U_{11} \left(\frac{V_{ub}V_{ts}}{V_{us}} - V_{tb} \right)   & U_{12} \left(\frac{V_{ub}V_{ts}}{V_{us}} - V_{tb} \right)     & 0   
\end{array}  
\right) \,, \\
\nonumber
\\
\label{eq:unu2}
\lambda_{u\nu}^{[e3B]} &=
\lambda_{be} \left(
\begin{array}{ccc}
U_{11} \left(\frac{V_{us}V_{cb}}{V_{cs}} - V_{ub} \right)   & U_{12} \left(\frac{V_{us}V_{cb}}{V_{cs}} - V_{ub} \right)  & 0 \\
0 & 0  & 0    \\
U_{11}\left(\frac{V_{cb}V_{ts}}{V_{cs}} - V_{tb} \right)  & U_{12} \left(\frac{V_{cb}V_{ts}}{V_{cs}} - V_{tb} \right)     & 0   
\end{array}
\right)  \, ,  \\
\nonumber
\\
\label{eq:unu3}
\lambda_{u\nu}^{[e3C]}  &=
\lambda_{be} \left(
\begin{array}{ccc}
U_{11} \left(\frac{V_{us}V_{tb}}{V_{ts}} - V_{ub} \right)   & U_{12} \left(\frac{V_{us}V_{tb}}{V_{ts}} - V_{ub} \right)  & 0 \\
U_{11}\left(\frac{V_{cs}V_{tb}}{V_{ts}} - V_{cb} \right)  & U_{12} \left(\frac{V_{cs}V_{tb}}{V_{ts}} - V_{cb} \right)     & 0   \\
0 & 0 & 0
\end{array}
\right)  \, , 
\end{align}
with $U_{13}$ already set to zero, and the three different matrices corresponding to the viable solutions of \eqref{eq:ulconstraint}.  These couplings are allowed by $T_{u,\nu}$ when $\alpha_{\nu} = \beta_{\nu} = - \beta_{u} = - \gamma_{u}$, $\alpha_{\nu} = \beta_{\nu} = - \alpha_{u} = - \gamma_{u}$ or $\alpha_{\nu} = \beta_{\nu} = - \alpha_{u} = - \beta_{u}$ respectively.  Observe that \eqref{eq:unu1}-\eqref{eq:unu3} do not permit isolation patterns for $\lambda_{u\nu}$, as this would force either $\lambda_{be}$, $U_{1i}$, or the special combinations of $V_{ij}$ seen in \eqref{eq:unu1}-\eqref{eq:unu3} to be zero, and none of these options are phenomenologically acceptable or interesting.

We therefore conclude that, when $\lambda_{dl}$ is of isolation pattern form and experimental data are considered, there are only three sets of viable couplings allowed by weak $SU(2)$ and residual flavour symmetries $T_{u,d,l,\nu}$:
\begin{equation}
\label{eq:finalisolation}
\lbrace \lambda_{dl},\lambda_{d\nu},\lambda_{ul}, \lambda_{u\nu} \rbrace \in
\begin{cases}
\lambda_{QL}^{e3A} \equiv \lbrace \lambda_{dl}^{[e3A]},\lambda_{d\nu}^{[e3A]},\lambda_{ul}^{[e3A]},\lambda_{u\nu}^{[e3A]} \rbrace  \\
\\
\lambda_{QL}^{e3B} \equiv \lbrace \lambda_{dl}^{[e3B]},\lambda_{d\nu}^{[e3B]},\lambda_{ul}^{[e3B]},\lambda_{u\nu}^{[e3B]}  \rbrace
\\
\\
\lambda_{QL}^{e3C} \equiv \lbrace \lambda_{dl}^{[e3C]},\lambda_{d\nu}^{[e3C]},\lambda_{ul}^{[e3C]},\lambda_{u\nu}^{[e3C]}  \rbrace
\end{cases}
\end{equation}
The associated phase constraints for each of these models are summarized in Table \ref{tab:finalphases} and all of the explicit matrices are given in Appendix \ref{sec:A}, where one can observe that each coupling depends on only one degree of freedom. Note that the leptoquark Yukawa couplings can have CP violation, but it depends exclusively on the CP-violating phases of the CKM and PMNS. Beyond CKM and PMNS elements, the couplings of the leptoquarks involved in each solution depend only on a single complex parameter (e.g. $\lambda_{be}$), whose phase can be absorbed by a rephasing of the leptoquark field.
Another important observation is that entries in the first and second row are suppressed by CKM combinations of order $\lambda^3$ and $\lambda^2$, respectively. In this they demonstrate a Froggatt-Nielsen type hierarchy in the quark sector of the leptoquark couplings, a feature which appears in the following section for sets of couplings of type $A$ (no coupling to $u$ quarks) and of type $B$ (no coupling to $c$ quarks). The sole exception is in the $[e3C]$ pattern which is of type $C$ (no coupling to $t$ quarks) which appears only in the isolation patterns: the coupling to $u$ quarks is still suppressed by $\lambda$ with respect to the coupling to $c$ quarks, however the coupling to $s$ quarks is stronger than to $b$ quarks.

This result is remarkably restrictive (and predictive).  Note also that, since in both the up and down sectors we can only resolve two generations, we should not expect to be able to predict the full three-generation CKM mixing within the confines of our strict residual symmetry approach.  On the other hand, three generation leptonic mixing may still be viable (up to the experimental caveat regarding $U^{13}_{PMNS}$ mentioned above), because the restriction $\alpha_{\nu} = \beta_{\nu}$ remains consistent with the Klein symmetry of the Majorana neutrino mass term!

\subsection{Two-columned Patterns: $\lambda_{dl} = \lambda_{dl}^{[e\mu, e \tau, \mu \tau]}$}
\label{sec:2column}
We now move to the case where $\lambda_{dl} = \lambda^{[e\mu,\mu\tau,e\tau]}$, i.e. where it takes a two-columned pattern due to the combined phase constraints of $T_{d,l}$, and in particular take the $\lambda^{[e\mu]}$ pattern for $\lambda_{dl}$ as a starting point.   The SU(2) prediction for the $\lambda_{d \nu}$ is then given by
\begin{equation}
\label{eq:dnu2column}
\lambda_{d\nu} =
\frac{1}{\sqrt{2}} \left(
\begin{array}{ccc}
0 & 0 & 0 \\
U_{11} \lambda_{se} +  U_{21} \lambda_{s\mu} & U_{12} \lambda_{se} +  U_{22} \lambda_{s\mu}  & U_{13} \lambda_{se} +  U_{23} \lambda_{s\mu}   \\
U_{11} \lambda_{be} +  U_{21} \lambda_{b\mu}  & U_{12} \lambda_{be} +  U_{22} \lambda_{b\mu} & U_{13} \lambda_{be} +  U_{23} \lambda_{b\mu} 
\end{array}
\right),
\end{equation}
where we have taken $\lambda_{de} = \lambda_{d\mu} = 0$ as required.  One immediately observes that an isolation pattern is not allowed for this coupling, as it would require at least two of the following equalities with PMNS mixing elements to be met:
\begin{equation}
\label{eq:2cemuSU2}
\vert \frac{U_{2 i}}{U_{1i}}\vert \overset{!}{=} \vert \frac{\lambda_{j e}}{\lambda_{j\mu}} \vert, \,\,\,\,\, \text{with} \,\,\,\,\, i \in \lbrace 1,2,3 \rbrace,\,\,\,\,\, j \in \lbrace s, b\rbrace ,
\end{equation}
with the $i$ sourced by the column of \eqref{eq:dnu2column} and where the leptoquark Yukawa couplings on the RHS are bound to the same experimental interval.  None of the NuFit bounds for the LHS of \eqref{eq:2cemuSU2} overlap, indicating that only one column in \eqref{eq:dnu2column} can be set to zero. As $i=2$ gives $\vert \frac{U_{22}}{U_{12}}\vert>1$ and $i=3$ gives $\vert \frac{U_{23}}{U_{13}} \vert >1$,  and we have from the geometric bound (\ref{eq:geobound}) that $\vert \frac{\lambda_{j e}}{\lambda_{j\mu}} \vert = \vert \kappa_e \vert \lesssim 0.71$, the only remaining solution to \eqref{eq:2cemuSU2} is given by $i = 1$, which forces this coupling into a two column pattern with non-zero entries in the second and third columns. This corresponds to $\beta_\nu = \gamma_\nu = - \beta_{d} = - \gamma_{d}$ and the following replacements for its couplings:
\begin{equation}
\label{eq:dnu2cconstraint}
\lambda_{se} = - \lambda_{s \mu} \frac{U_{21}}{U_{11}}, \,\,\,\,\,\,\,\,\,\, \lambda_{be} = - \lambda_{b \mu} \frac{U_{21}}{U_{11}},
\end{equation}  
such that the $d-\nu$ matrix becomes
\begin{equation}
\label{eq:dnuemu}
\lambda_{d\nu}^{[e \mu 1]} =
\frac{1}{\sqrt{2}} \left(
\begin{array}{ccc}
0 & 0 & 0 \\
0 & \left(-\frac{U_{12}U_{21}}{U_{11}} + U_{22} \right) \lambda_{s\mu}  & \left(-\frac{U_{13}U_{21}}{U_{11}} + U_{23} \right) \lambda_{s\mu}    \\
0 & \left(-\frac{U_{12}U_{21}}{U_{11}} + U_{22} \right) \lambda_{b\mu}  &  \left(-\frac{U_{13}U_{21}}{U_{11}} + U_{23} \right) \lambda_{b\mu} 
\end{array}
\right)
\end{equation}
and where \eqref{eq:dnu2cconstraint} also obviously applies to $\lambda_{dl}^{[e\mu]}$.

Continuing to the $u-l$ coupling, from $\lambda_{dl}^{[e\mu]}$ we set $\lambda_{d(e,\mu)} = 0$, and applying \eqref{eq:dnu2cconstraint} from above, we find
\begin{equation}
\label{eq:ul2column}
\lambda_{ul} =
\frac{1}{\sqrt{2}} \left(
\begin{array}{ccc}
-\frac{U_{21}}{U_{11}}\left(V_{ub} \lambda_{b\mu} +  V_{us} \lambda_{s\mu}\right)  & V_{ub} \lambda_{b\mu} + V_{us} \lambda_{s \mu} & 0 \\
-\frac{U_{21}}{U_{11}}\left(V_{cb} \lambda_{b\mu} +  V_{cs} \lambda_{s\mu}\right) & V_{cb} \lambda_{b\mu} + V_{cs} \lambda_{s \mu}  & 0  \\
-\frac{U_{21}}{U_{11}}\left(V_{tb} \lambda_{b\mu} +  V_{ts} \lambda_{s\mu}\right) & V_{tb} \lambda_{b\mu} + V_{ts} \lambda_{s \mu}& 0
\end{array}
\right) \, .
\end{equation}
At least one row must be set to zero in order to satisfy our phase restrictions, and as the ratio of PMNS elements in the first column is not consistent with zero, we are left demanding
\begin{equation}
\label{eq:2cemuSU2ul}
\vert \frac{V_{ib}}{V_{is}} \vert \overset{!}{=} \vert \frac{\lambda_{s \mu}}{\lambda_{b \mu}} \vert, \,\,\,\,\, \text{with} \,\,\,\,\, i \in \lbrace u,c,t \rbrace .
\end{equation}
For $i=t$, this condition leads to $\vert \frac{V_{tb}}{V_{ts}} \vert \gg 1$, which is not consistent with $\vert \frac{\lambda_{s \mu}}{\lambda_{b \mu}} \vert =\vert \rho \vert \lesssim 1$. Thus, the condition can only be met for $i = \lbrace u,c \rbrace$, which allows two patterns for $\lambda_{ul}$, one with the first row set to zero, called $\lambda_{ul}^{[e\mu A]}$, and the other with the second row set to zero, called $\lambda_{ul}^{[e\mu B]}$:
\begin{align}
\label{eq:ulemuA}
\lambda_{ul}^{[e \mu A]} &=
\frac{ \lambda_{b\mu}}{\sqrt{2}} \left(
\begin{array}{ccc}
0  & 0 & 0 \\
\frac{U_{21}}{U_{11}}\left(\frac{V_{ub}V_{cs}}{V_{us}} - V_{cb} \right)  & \left(-\frac{V_{ub}V_{cs}}{V_{us}} + V_{cb} \right) & 0  \\
\frac{U_{21}}{U_{11}}\left(\frac{V_{ub}V_{ts}}{V_{us}} - V_{tb} \right)&  \left(-\frac{V_{ub}V_{ts}}{V_{us}} + V_{tb} \right)  & 0
\end{array}
\right) \, ,\\
\nonumber
\\
\lambda_{ul}^{[e \mu B]} &=
\frac{ \lambda_{b\mu}}{\sqrt{2}} \left(
\begin{array}{ccc}
\frac{U_{21}}{U_{11}}\left(\frac{V_{us}V_{cb}}{V_{cs}} - V_{ub} \right) & \left(-\frac{V_{us}V_{cb}}{V_{cs}} + V_{ub} \right) & 0  \\
0  & 0 & 0 \\
\frac{U_{21}}{U_{11}}\left(\frac{V_{cb}V_{ts}}{V_{cs}} - V_{tb} \right)  &  \left(-\frac{V_{cb}V_{ts}}{V_{cs}} + V_{tb} \right)  & 0
\end{array}
\right) \, .
\end{align}
For $\lambda_{ul}^{[e\mu A]}$, $\alpha_{l} = \beta_{l} = -\beta_{u} = - \gamma_{u}$, whereas for $\lambda_{ul}^{[e\mu B]}$  $\alpha_{l} = \beta_{l} = -\alpha_{u} = - \gamma_{u}$.


Finally we consider the $u-\nu$ term. At this stage we have already fixed phase constraints in all four fermionic sectors, so we need to check if the resulting structures are consistent with non-zero couplings. Applying all relevant constraints sourced from the $d-l$, $d-\nu$, and $u-l$ couplings as well as $SU(2)$ relationships, we find that for $\lambda_{dl}^{[e\mu 1(A,B)]}$ the couplings are automatically found in symmetric forms:
\begin{align}
\label{eq:unu2column}
\lambda_{u\nu}^{[e\mu 1A]} &=
\lambda_{b\mu} \left(
\begin{array}{ccc}
0  & 0 & 0 \\
0 & \left( \frac{U_{12} U_{21}}{U_{11}} - U_{22} \right) \left(-\frac{V_{ub}V_{cs}}{V_{us}} +V_{cb} \right)  &  \left( \frac{U_{13} U_{21}}{U_{11}} - U_{23} \right) \left(-\frac{V_{ub}V_{cs} }{V_{us}} +V_{cb} \right)  \\
0 & \left( \frac{U_{12} U_{21}}{U_{11}} - U_{22} \right) \left(-\frac{V_{ub}V_{ts}}{V_{us}} +V_{tb} \right) & \left( \frac{U_{13} U_{21}}{U_{11}} - U_{23} \right) \left(-\frac{V_{ub}V_{ts} }{V_{us}} +V_{tb} \right) 
\end{array}
\right) \, ,
\\
\nonumber
\\
\label{eq:unu2columnB}
\lambda_{u\nu}^{[e\mu 1B]} &=
\lambda_{b\mu} \left(
\begin{array}{ccc}
0 & \left( \frac{U_{12} U_{21}}{U_{11}} - U_{22} \right) \left(-\frac{V_{cb}V_{us}}{V_{cs}} + V_{ub} \right) &  \left( \frac{U_{13} U_{21}}{U_{11}} - U_{23} \right) \left(-\frac{V_{us}V_{cb}}{V_{cs}} + V_{ub} \right)  \\
0  & 0 & 0 \\
0 &\left(\frac{U_{12} U_{21} }{U_{11}} - U_{22} \right) \left(-\frac{V_{cb}V_{ts}}{V_{cs}} + V_{tb} \right) & \left( \frac{U_{13} U_{21}}{U_{11}} - U_{23} \right) \left(- \frac{V_{cb}V_{ts}}{V_{cs}} + V_{tb}  \right)\end{array}
\right) \, .
\end{align}

Starting from $\lambda_{dl}^{[e\mu]}$ we arrive at two possible solutions. We now denote these as
\begin{align}
\lambda_{QL}^{e\mu 1A}  &\equiv \lbrace \lambda_{dl}^{[e\mu 1A]},\lambda_{d\nu}^{[e\mu 1A]},\lambda_{ul}^{[e\mu 1A]},\lambda_{u\nu}^{[e\mu 1A]} \rbrace , \\
\lambda_{QL}^{e\mu 1B} &\equiv \lbrace \lambda_{dl}^{[e\mu 1B]},\lambda_{d\nu}^{[e\mu 1B]},\lambda_{ul}^{[e\mu 1B]},\lambda_{u\nu}^{[e\mu 1B]}  \rbrace ,
\end{align}
with the superscript `$1$' denoting the vanishing column in couplings to neutrinos and the $A,B$ denoting whether it is the first or second row that vanishes in couplings to up quarks.
Note that although very similar, $\lambda_{dl}^{[e\mu 1A]} \neq \lambda_{dl}^{[e\mu 1B]}$ and $\lambda_{d\nu}^{[e\mu 1A]} \neq \lambda_{d\nu}^{[e\mu 1B]}$ as they have different CKM elements, as can also be seen in Appendix \ref{sec:A} where we list all coupling matrices for the possible solutions. As in the isolation patterns, we see the column-dependent PMNS modulations, and the very interesting row-dependent suppressions that are in the style of Froggatt-Nielsen symmetries of $\lambda^3$ for the first row and $\lambda^2$ for the second row.  Also, as we have already mentioned for the isolation patterns, CP violation depends solely on the CP-violating phases of the CKM and PMNS, as the single parameter (e.g. $\lambda_{b\mu}$) that appears can be made real without loss of generality through an appropriate rephasing of the leptoquark field.

The derivation of the couplings permitted when $\lambda_{dl} = \lambda_{dl}^{[e\tau]}$ and $\lambda_{dl} = \lambda_{dl}^{[\mu\tau]}$ follow in direct analogue to the $\lambda_{dl}^{[e\mu]}$ case, so we do not show them explicitly here.  When $\lambda_{dl} = \lambda_{dl}^{[e\tau]}$, noting again the underlying assumption that the hierarchy in the leptoquark couplings follows that of the charged leptons (in this case, the $e$ column has smaller entries than the $\tau$ column), the $\kappa_e$ bound coming from (\ref{eq:geobound}) combines with the allowed ranges for PMNS entries and leaves only the solution where the first column of $d-\nu$ vanishes ($i=1$). The resulting patterns are $\lambda_{QL}^{e\tau 1A}$, $\lambda_{QL}^{e\tau 1B}$.  On the other hand, when $\lambda_{dl} = \lambda_{dl}^{[\mu\tau]}$, the analogous (\ref{eq:geobound}) constraint for $\kappa_\mu$ (where here the $\mu$ column has smaller entries than the $\tau$ column) combines with the allowed ranges of PMNS entries and allows only $i=1$ solutions for $d-\nu$. This results in two sets of couplings, $\lambda_{QL}^{\mu\tau 1A}$, $\lambda_{QL}^{\mu\tau 1B}$. Here it is important to note that the numerical values shown in Appendix \ref{sec:A} are meant for illustration and correspond only to central values of the allowed ranges, which is why (counterintuitively) the numerical values shown for these patterns in particular do not fulfill the $\kappa_\mu \lesssim 0.71$ constraint. If the experimentally allowed ranges for the respective PMNS entries narrow down and remain close to the current central values, these patterns would be ruled out.

In total, assuming leptoquark hierarchies following those of the down quark and charged leptons, we have found that there are six unique sets of leptoquark couplings generated from the original two-columned patterns of $\lambda_{dl}$.  We denote these patterns as 
\begin{equation}
\label{eq:finalunique}
 \lambda_{dl,d\nu,ul,u\nu} \in \left\{ \begin{alignedat}{2}
\lambda_{QL}^{e\mu 1A}  &\equiv \lbrace \lambda_{dl}^{[e\mu 1A]},\lambda_{d\nu}^{[e\mu 1A]},\lambda_{ul}^{[e\mu 1A]},\lambda_{u\nu}^{[e\mu 1A]} \rbrace, \,\,\,\,\,\,\,\,\,\,
 &&\lambda_{QL}^{e\mu 1B} \equiv \lbrace \lambda_{dl}^{[e\mu 1B]},\lambda_{d\nu}^{[e\mu 1B]},\lambda_{ul}^{[e\mu 1B]},\lambda_{u\nu}^{[e\mu 1B]}  \rbrace   \\
\lambda_{QL}^{e\tau 1A} &\equiv \lbrace \lambda_{dl}^{[e\tau 1A]},\lambda_{d\nu}^{[e\tau 1A]},\lambda_{ul}^{[e\tau 1A]},\lambda_{u\nu}^{[e\tau 1A]} \rbrace,  \,\,\,\,\,\,\,\,\,\,
&&\lambda_{QL}^{e\tau 1B} \equiv \lbrace \lambda_{dl}^{[e\tau 1B]},\lambda_{d\nu}^{[e\tau 1B]},\lambda_{ul}^{[e\tau 1B]},\lambda_{u\nu}^{[e\tau 1B]}  \rbrace \\
 \lambda_{QL}^{\mu\tau 1A} &\equiv \lbrace \lambda_{dl}^{[\mu\tau 1A]},\lambda_{d\nu}^{[\mu\tau 1A]},\lambda_{ul}^{[\mu\tau 1A]},\lambda_{u\nu}^{[\mu \tau 1A]}  \rbrace,  \,\,\,\,\,\,\,\,\,\,
&&\lambda_{QL}^{\mu\tau 1B} \equiv \lbrace \lambda_{dl}^{[\mu\tau 1B]},\lambda_{d\nu}^{[\mu\tau 1B]},\lambda_{ul}^{[\mu\tau 1 B]},\lambda_{u\nu}^{[\mu\tau 1B]}  \rbrace \\
\end{alignedat}\right.
\end{equation}
and their associated phase and matrix element constraints can be found in Table \ref{tab:finalphases}.  All of the explicit matrices are given in Appendix \ref{sec:A} for easy reference.  \\

\subsection{Three-columned Patterns:  $\lambda_{dl} = \lambda_{dl}^{[e\mu1,1\mu\tau,e1\tau]}$}
\label{sec:threecolumn}

The final set of patterns allowed for $\lambda_{dl}$ is given by the three-columned matrices of \eqref{eq:yukemt}.  Applying \eqref{eq:LQrelations} to $\lambda_{dl}^{[e\mu1]}$ and first forming the $d-\nu$ coupling, one finds
\begin{equation}
\label{eq:dnu3column}
\lambda_{d\nu} =
\frac{1}{\sqrt{2}} \left(
\begin{array}{ccc}
U_{31} \lambda_{d\tau}  & U_{32} \lambda_{d\tau} & U_{33} \lambda_{d\tau} \\
 U_{11} \lambda_{se} + U_{21} \lambda_{s\mu} &  U_{12} \lambda_{se} + U_{22} \lambda_{s\mu} &  U_{13} \lambda_{se} + U_{23} \lambda_{s\mu}  \\
 U_{11} \lambda_{be} + U_{21} \lambda_{b\mu} &  U_{12} \lambda_{be} + U_{22} \lambda_{b\mu} &  U_{13} \lambda_{be} + U_{23} \lambda_{b\mu} 
\end{array}
\right)\,,
\end{equation}
from which one immediately concludes that the only viable option is to set $\lambda_{d\tau} \rightarrow 0$ (because there must be at least one zero in every row), which then reduces this pattern to that of \eqref{eq:dnu2column}, and therefore all of the associated constraints derived for that matrix also hold for \eqref{eq:dnu3column}.  Indeed, the constraint $\lambda_{d\tau} \rightarrow 0$ simultaneously reduces the three-columned $\lambda_{dl}$ to its two-columned cousin.  In fact, had we instead considered $\lambda_{dl}^{[1\mu\tau]}$ or $\lambda_{dl}^{[e1\tau]}$, we would have analogously found that $\lambda_{de} \rightarrow 0$ or $\lambda_{d\mu} \rightarrow 0$, respectively, such that those matrices also reduce to their two-columned `special-case'.  

\emph{We are therefore led to conclude that, modulo the $U_{13}$ caveat discussed in Section \ref{sec:isolation}, the nine unique sets of leptoquark Yukawa couplings allowed are those given in Table \ref{tab:finalphases}, when considering the normal hierarchy of leptoquark couplings.}

\begin{table} [t]
\centering
\small
{\renewcommand{\arraystretch}{1.65}
\begin{tabular}{|c|c||c|c|c|}
\hline
 \multicolumn{2}{|c||}{$\lambda_{QL}$} & \multicolumn{2}{c|}{Phase Equalities}  & Notes \\
\hline
\hline
\parbox[t]{2mm}{\multirow{9}{*}{\rotatebox[origin=c]{90}{Isolation Patterns}}} & \multirow{3}{*}{$\lambda_{QL}^{e3A}$} & $\Delta_{3}$ &$\lbrace$ $\beta_{d}$, $\gamma_{d}$, $-\alpha_{\nu}$, $-\beta_{\nu}$, $-\alpha_{l}$, $\beta_{u}$, $\gamma_{u}$ $\rbrace$ & \multirow{3}{*}{$U_{13} = 0$, $\lambda_{se} = - \lambda_{be} \frac{V_{ub}}{V_{us}}$}\\
\cline{3-4}
& & $\Delta_{3}^{\mu}$  & $\lbrace$ $\beta_{d}$, $\gamma_{d}$, $\alpha_{\nu}$, $\beta_{\nu}$, $\alpha_{l}$, $\beta_{u}$, $\gamma_{u}$ $\rbrace$ & \\
\cline{3-4}
& & $\Delta_{1}^{\mu}$ & $\lbrace$ $\beta_{d}, \gamma_{d}, \alpha_{l}$ $\rbrace$ $\lbrace$ $\alpha_{\nu}, \beta_{\nu}, \beta_{u}, \gamma_{u}$ $\rbrace$ & \\
\cline{2-5}
& \multirow{3}{*}{$\lambda_{QL}^{e3B}$} & $\Delta_{3}$ & $\lbrace$ $\beta_{d}$, $\gamma_{d}$, $-\alpha_{\nu}$, $-\beta_{\nu}$, $-\alpha_{l}$, $\alpha_{u}$, $\gamma_{u}$ $\rbrace$ & \multirow{3}{*}{$U_{13} = 0$, $\lambda_{se} = - \lambda_{be} \frac{V_{cb}}{V_{cs}}$} \\
\cline{3-4}
& & $\Delta_{3}^{\mu}$  & $\lbrace$ $\beta_{d}$, $\gamma_{d}$, $\alpha_{\nu}$, $\beta_{\nu}$, $\alpha_{l}$, $\alpha_{u}$, $\gamma_{u}$ $\rbrace$ & \\
\cline{3-4}
& & $\Delta_{1}^{\mu}$ & $\lbrace$ $\beta_{d}, \gamma_{d}, \alpha_{l}$ $\rbrace$ $\lbrace$ $\alpha_{\nu}, \beta_{\nu}, \alpha_{u}, \gamma_{u}$ $\rbrace$ & \\
\cline{2-5}
& \multirow{3}{*}{$\lambda_{QL}^{e3C}$} & $\Delta_{3}$ & $\lbrace$ $\beta_{d}$, $\gamma_{d}$, $-\alpha_{\nu}$, $-\beta_{\nu}$, $-\alpha_{l}$, $\alpha_{u}$, $\beta_{u}$ $\rbrace$ & \multirow{3}{*}{$U_{13} = 0$, $\lambda_{se} = - \lambda_{be} \frac{V_{tb}}{V_{ts}}$} \\
\cline{3-4}
& & $\Delta_{3}^{\mu}$  & $\lbrace$ $\beta_{d}$, $\gamma_{d}$, $\alpha_{\nu}$, $\beta_{\nu}$, $\alpha_{l}$, $\alpha_{u}$, $\beta_{u}$ $\rbrace$ & \\
\cline{3-4}
& & $\Delta_{1}^{\mu}$ & $\lbrace$ $\beta_{d}, \gamma_{d}, \alpha_{l}$ $\rbrace$ $\lbrace$ $\alpha_{\nu}, \beta_{\nu}, \alpha_{u}, \beta_{u}$ $\rbrace$ & \\
\cline{2-5}
\hline
\hline
\parbox[t]{2mm}{\multirow{6}{*}{\rotatebox[origin=c]{90}{$e-\mu$ \,\, Patterns}}} & \multirow{3}{*}{$\lambda_{QL}^{e\mu 1A}$}  &  $\Delta_{3}$ &
$\lbrace$ $\beta_{d}$, $\gamma_{d}$, $-\beta_{\nu}$,  $-\gamma_{\nu}$,  $-\alpha_{l}$,  $-\beta_{l}$,  $\beta_{u}$,  $\gamma_{u}$ $\rbrace$ 
& \multirow{3}{*}{$\lambda_{se} = - \lambda_{s \mu} \frac{U_{21}}{U_{11}}$, $\lambda_{be} = - \lambda_{b \mu} \frac{U_{21}}{U_{11}}$ , $\lambda_{s\mu} = - \lambda_{b \mu}  \frac{V_{ub}}{V_{us} }$} \\
\cline{3-4}
& & $\Delta_{3}^{\mu}$  & $\lbrace$ $\beta_{d}$, $\gamma_{d}$, $\beta_{\nu}$,  $\gamma_{\nu}$,  $\alpha_{l}$,  $\beta_{l}$,  $\beta_{u}$,  $\gamma_{u}$ $\rbrace$ & \\
\cline{3-4}
& & $\Delta_{1}^{\mu}$ & $\lbrace$ $\beta_{d}, \gamma_{d}, \alpha_{l}, \beta_{l}$ $\rbrace$ $\lbrace$ $\beta_{\nu}, \gamma_{\nu}, \beta_{u}, \gamma_{u}$ $\rbrace$ & \\
\cline{2-5}
&\multirow{3}{*}{$ \lambda_{QL}^{e\mu 1B}$} & $\Delta_{3}$ &
$\lbrace$ $\beta_{d}$, $\gamma_{d}$, $-\beta_{\nu}$,  $-\gamma_{\nu}$,  $-\alpha_{l}$,  $-\beta_{l}$,  $\alpha_{u}$,  $\gamma_{u}$ $\rbrace$ 
& \multirow{3}{*}{$\lambda_{se} = - \lambda_{s \mu} \frac{U_{21}}{U_{11}}$, $\lambda_{be} = - \lambda_{b \mu} \frac{U_{21}}{U_{11}}$ , $\lambda_{s\mu} = - \lambda_{b \mu}  \frac{V_{cb}}{V_{cs} }$} \\
\cline{3-4}
& & $\Delta_{3}^{\mu}$  & $\lbrace$ $\beta_{d}$, $\gamma_{d}$, $\beta_{\nu}$,  $\gamma_{\nu}$,  $\alpha_{l}$,  $\beta_{l}$,  $\alpha_{u}$,  $\gamma_{u}$ $\rbrace$  & \\
\cline{3-4}
& & $\Delta_{1}^{\mu}$ & $\lbrace$ $\beta_{d}, \gamma_{d}, \alpha_{l}, \beta_{l} $ $\rbrace$ $\lbrace$ $\beta_{\nu}, \gamma_{\nu}, \alpha_{u}, \gamma_{u}$ $\rbrace$ & \\
\cline{2-5}
\hline
\hline
\parbox[t]{2mm}{\multirow{6}{*}{\rotatebox[origin=c]{90}{$e-\tau $ \,\, Patterns}}} &\multirow{3}{*}{$\lambda_{QL}^{e\tau 1A}$} & $\Delta_{3}$ &
$\lbrace$ $\beta_{d}$, $\gamma_{d}$, $-\beta_{\nu}$,  $-\gamma_{\nu}$,  $-\alpha_{l}$,  $-\gamma_{l}$,  $\beta_{u}$,  $\gamma_{u}$ $\rbrace$ 
& \multirow{3}{*}{$\lambda_{se} = - \lambda_{s \tau} \frac{U_{31}}{U_{11}}$, $\lambda_{be} = - \lambda_{b \tau} \frac{U_{31}}{U_{11}}$, $\lambda_{s\tau} = - \lambda_{b \tau} \frac{V_{ub}}{V_{us}}$} \\
\cline{3-4}
& & $\Delta_{3}^{\mu}$  & $\lbrace$ $\beta_{d}$, $\gamma_{d}$, $\beta_{\nu}$,  $\gamma_{\nu}$,  $\alpha_{l}$,  $\gamma_{l}$,  $\beta_{u}$,  $\gamma_{u}$ $\rbrace$ & \\
\cline{3-4}
& & $\Delta_{1}^{\mu}$ & $\lbrace$ $\beta_{d}, \gamma_{d}, \alpha_{l}, \gamma_{l}$ $\rbrace$ $\lbrace$ $\beta_{\nu}, \gamma_{\nu}, \beta_{u}, \gamma_{u}$ $\rbrace$ & \\
\cline{2-5}
&\multirow{3}{*}{$\lambda_{QL}^{e\tau 1B}$} & $\Delta_{3}$ &
$\lbrace$ $\beta_{d}$, $\gamma_{d}$, $-\beta_{\nu}$,  $-\gamma_{\nu}$,  $-\alpha_{l}$,  $-\gamma_{l}$,  $\alpha_{u}$,  $\gamma_{u}$ $\rbrace$ 
& \multirow{3}{*}{$\lambda_{se} = - \lambda_{s \tau} \frac{U_{31}}{U_{11}}$, $\lambda_{be} = - \lambda_{b \tau} \frac{U_{31}}{U_{11}}$, $\lambda_{s\tau} = - \lambda_{b \tau} \frac{V_{cb}}{V_{cs}}$} \\
\cline{3-4}
& & $\Delta_{3}^{\mu}$  & $\lbrace$ $\beta_{d}$, $\gamma_{d}$, $\beta_{\nu}$,  $\gamma_{\nu}$,  $\alpha_{l}$,  $\gamma_{l}$,  $\alpha_{u}$,  $\gamma_{u}$ $\rbrace$ & \\
\cline{3-4}
& & $\Delta_{1}^{\mu}$ & $\lbrace$ $\beta_{d}, \gamma_{d}, \alpha_{l}, \gamma_{l}$ $\rbrace$ $\lbrace$ $\beta_{\nu}, \gamma_{\nu}, \alpha_{u}, \gamma_{u}$ $\rbrace$ & \\
\hline
\hline
\parbox[t]{2mm}{\multirow{6}{*}{\rotatebox[origin=c]{90}{$\mu-\tau $ \,\, Patterns}}} & \multirow{3}{*}{$\lambda_{QL}^{\mu\tau 1A}$} & $\Delta_{3}$ &
$\lbrace$ $\beta_{d}$, $\gamma_{d}$, $-\beta_{\nu}$,  $-\gamma_{\nu}$,  $-\beta_{l}$,  $-\gamma_{l}$,  $\beta_{u}$,  $\gamma_{u}$ $\rbrace$ 
& \multirow{3}{*}{$\lambda_{s\mu} = - \lambda_{s \tau} \frac{U_{31}}{U_{21}}$, $\lambda_{b\mu} = - \lambda_{b \tau} \frac{U_{31}}{U_{21}}$, $ \lambda_{s\tau} = -\lambda_{b \tau}\frac{V_{ub}}{V_{us}} $} \\
\cline{3-4}
& & $\Delta_{3}^{\mu}$  & $\lbrace$ $\beta_{d}$, $\gamma_{d}$, $\beta_{\nu}$,  $\gamma_{\nu}$,  $\beta_{l}$,  $\gamma_{l}$,  $\beta_{u}$,  $\gamma_{u}$ $\rbrace$ & \\
\cline{3-4}
& & $\Delta_{1}^{\mu}$ & $\lbrace$ $\beta_{d}, \gamma_{d}, \beta_{l}, \gamma_{l}$ $\rbrace$ $\lbrace$ $\beta_{\nu}, \gamma_{\nu}, \beta_{u}, \gamma_{u}$ $\rbrace$ & \\
\cline{2-5}
&\multirow{3}{*}{$\lambda_{QL}^{\mu\tau 1B}$} & $\Delta_{3}$ &
$\lbrace$ $\beta_{d}$, $\gamma_{d}$, $-\beta_{\nu}$,  $-\gamma_{\nu}$,  $-\beta_{l}$,  $-\gamma_{l}$,  $\alpha_{u}$,  $\gamma_{u}$ $\rbrace$ 
& \multirow{3}{*}{$\lambda_{s\mu} = - \lambda_{s \tau} \frac{U_{31}}{U_{21}}$, $\lambda_{b\mu} = - \lambda_{b \tau} \frac{U_{31}}{U_{21}}$, $ \lambda_{s\tau} = -\lambda_{b \tau}\frac{V_{cb}}{V_{cs}} $} \\
\cline{3-4}
& & $\Delta_{3}^{\mu}$  & $\lbrace$ $\beta_{d}$, $\gamma_{d}$, $\beta_{\nu}$,  $\gamma_{\nu}$,  $\beta_{l}$,  $\gamma_{l}$,  $\alpha_{u}$,  $\gamma_{u}$ $\rbrace$ & \\
\cline{3-4}
& & $\Delta_{1}^{\mu}$ & $\lbrace$ $\beta_{d}, \gamma_{d}, \beta_{l}, \gamma_{l}$ $\rbrace$ $\lbrace$ $\beta_{\nu}, \gamma_{\nu}, \alpha_{u}, \gamma_{u}$ $\rbrace$ & \\
\cline{2-5}
\hline
\end{tabular}}
\caption{Simplified models of flavourful leptoquarks determined after symmetry and experimental constraints are applied.  The second column gives the set of couplings as defined in \eqref{eq:finalisolation} and \eqref{eq:finalunique}.  The fourth column gives all phases that must be set equal to one another for the scalar and vector cases.  Finally, the fifth column gives the relationships between the matrix elements of the original $d-l$ coupling term.  NOTE:  For the vectors $\Delta_{(1,3)}^{\mu}$, replace $V_{ij} \rightarrow V^{\star}_{ij}$.}
\label{tab:finalphases}
\end{table}

\subsection{Implications for Proton Decay}
\label{sec:QQ}

We now discuss the implications of the residual symmetries $T_u$ and $T_d$ for diquark operators allowing proton decay in the scalar triplet case. In general we have
 \begin{align}
 \label{eq:LQdiquarkUU}
\lambda_{uu}:&& \,\,\,\,\,\,\,\,\,\,
    \left(
\begin{array}{ccc}
0 & e^{i(\alpha_{u} + \beta_{u})}\,\lambda_{uc^{\prime}} & e^{i(\alpha_{u} + \gamma_{u})}\,\lambda_{ut^{\prime}}  \\
-e^{i(\alpha_{u} + \beta_{u})}\,\lambda_{uc^{\prime}}  &  0  & e^{i(\beta_{u} + \gamma_{u})}\,\lambda_{ct^{\prime}}     \\
-e^{i(\alpha_{u} + \gamma_{u})}\,\lambda_{ut^{\prime}}  & -e^{i(\beta_{u} + \gamma_{u})}\,\lambda_{ct^{\prime}}   & 0 
\end{array}
\right)
  &\overset{!}{=}
  \left(
\begin{array}{ccc}
0 & \lambda_{uc^{\prime}} & \lambda_{ut^{\prime}}  \\
-\lambda_{uc^{\prime}}  &  0  & \lambda_{ct^{\prime}}     \\
-\lambda_{ut^{\prime}}  & -\lambda_{ct^{\prime}}   & 0 
\end{array}
\right)\\
\nonumber 
\\
 \label{eq:LQdiquarkDD}
\lambda_{dd}:&& \,\,\,\,\,\,\,\,\,\,
    \left(
\begin{array}{ccc}
0 & e^{i(\alpha_{d}+\beta_{d})}\,\lambda_{ds^{\prime}} & e^{i(\alpha_{d}+\gamma_{d})}\,\lambda_{db^{\prime}}  \\
-e^{i(\alpha_{d}+\beta_{d})}\,\lambda_{ds^{\prime}}  &  0  & e^{i(\beta_{d}+\gamma_{d})}\,\lambda_{sb^{\prime}}     \\
-e^{i(\alpha_{d}+\gamma_{d})}\,\lambda_{db^{\prime}}  & -e^{i(\beta_{d}+\gamma_{d})}\,\lambda_{sb^{\prime}}   & 0 
\end{array}
\right)
  &\overset{!}{=}
  \left(
\begin{array}{ccc}
0 & \lambda_{ds^{\prime}} & \lambda_{db^{\prime}}  \\
-\lambda_{ds^{\prime}}  &  0  & \lambda_{sb^{\prime}}     \\
-\lambda_{db^{\prime}}  & -\lambda_{sb^{\prime}}   & 0 
\end{array}
\right)\\
\nonumber 
\\
\label{eq:LQdiquarkDU}
\lambda_{du}:&& \,\,\,\,\,\,\,\,\,\,
    \left(
\begin{array}{ccc}
e^{i(\alpha_{d}+\alpha_{u})}\, \lambda_{du^{\prime}}  & e^{i(\alpha_{d}+\beta_{u})}\, \lambda_{dc^{\prime}} &  e^{i(\alpha_{d}+\gamma_{u})}\,\lambda_{dt^{\prime}}  \\
e^{i(\beta_{d}+\alpha_{u})}\,\lambda_{su^{\prime}}  & e^{i(\beta_{d}+\beta_{u})}\, \lambda_{sc^{\prime}}  & e^{i(\beta_{d}+\gamma_{u})}\, \lambda_{st^{\prime}}     \\
e^{i(\gamma_{d}+\alpha_{u})}\,\lambda_{bu^{\prime}}  & e^{i(\gamma_{d}+\beta_{u})}\,\lambda_{bc^{\prime}}   &  e^{i(\gamma_{d}+\gamma_{u})}\, \lambda_{bt^{\prime}}
\end{array}
\right)
  &\overset{!}{=}
  \left(
\begin{array}{ccc}
\lambda_{du^{\prime}} & \lambda_{dc^{\prime}} & \lambda_{dt^{\prime}}  \\
\lambda_{su^{\prime}}  &  \lambda_{sc^{\prime}}   & \lambda_{st^{\prime}}     \\
\lambda_{bu^{\prime}}  & \lambda_{bc^{\prime}}   &  \lambda_{bt^{\prime}}  
\end{array}
\right) \\
\nonumber 
\\
\label{eq:LQdiquarkUD}
\lambda_{ud}:&& \,\,\,\,\,\,\,\,\,\,
    \left(
\begin{array}{ccc}
e^{i(\alpha_{d}+\alpha_{u})}\, \lambda_{ud^{\prime}}  & e^{i(\alpha_{d}+\beta_{u})}\, \lambda_{cd^{\prime}} &  e^{i(\alpha_{d}+\gamma_{u})}\,\lambda_{td^{\prime}}  \\
e^{i(\beta_{d}+\alpha_{u})}\,\lambda_{us^{\prime}}  & e^{i(\beta_{d}+\beta_{u})}\, \lambda_{cs^{\prime}}  & e^{i(\beta_{d}+\gamma_{u})}\, \lambda_{ts^{\prime}}     \\
e^{i(\gamma_{d}+\alpha_{u})}\,\lambda_{ub^{\prime}}  & e^{i(\gamma_{d}+\beta_{u})}\,\lambda_{cb^{\prime}}   &  e^{i(\gamma_{d}+\gamma_{u})}\, \lambda_{tb^{\prime}}
\end{array}
\right)
  &\overset{!}{=}
  \left(
\begin{array}{ccc}
 \lambda_{ud^{\prime}}  & \lambda_{cd^{\prime}} &  \lambda_{td^{\prime}}  \\
\lambda_{us^{\prime}}  &  \lambda_{cs^{\prime}}  &  \lambda_{ts^{\prime}}     \\
\lambda_{ub^{\prime}}  & \lambda_{cb^{\prime}}   &  \lambda_{tb^{\prime}}
\end{array}
\right) 
\end{align}
where the primes simply differentiate matrix elements from the overall coupling matrices, and where the anti-symmetric condition \eqref{eq:antisymm} has been applied in \eqref{eq:LQdiquarkUU}-\eqref{eq:LQdiquarkDD}.

Proton decay is sensitive to elements coupling to a first generation quark, and so in order to maintain stability we must forbid entries in the first row and column of \eqref{eq:LQdiquarkUU}-\eqref{eq:LQdiquarkUD}. From \eqref{eq:LQdiquarkUU}-\eqref{eq:LQdiquarkDD} we immediately see that non-vanishing entries are only possible when additional phase relations are realized within each residual symmetry generator.  For example, $\lambda_{uc^{\prime}}$ is only allowed if $\alpha_u = -\beta_u$, and $\lambda_{ds}$ similarly requires $\alpha_{d} = - \beta_{d}$. On the other hand, it is clear from \eqref{eq:LQdiquarkDU}-\eqref{eq:LQdiquarkUD} that dangerous non-zero entries require equalities of the type $\alpha_u=-\alpha_d$ (and so on) between the phases of $T_u$ and $T_d$.  Yet these observations are generic --- we have not applied any of the conclusions from Sections \ref{sec:isolation}-\ref{sec:threecolumn}.  Indeed, comparing to the phase equalities in the scalar triplet `solutions' listed in Table \ref{tab:finalphases}, we note that no dangerous couplings are \emph{required} from these relationships alone, and so we are free to align generic $T_u$ and $T_d$ phases so that they vanish.   In other words, not only are all of the phenomenologically acceptable patterns derived in Table \ref{tab:finalphases} a priori compatible with models that forbid proton decay, the residual symmetry mechanism can be further exploited to forbid the decay without additional model-building assumptions.

As a pedagogical example of this possibility, we subject the diquark coupling matrices in \eqref{eq:LQdiquarkUU}-\eqref{eq:LQdiquarkUD} to the phase equalities implied in $\lambda_{QL}^{[e3A]}$:
 \begin{align}
 \label{eq:LQ1diquarkUUb}
\lambda_{uu}:&& \,\,\,\,\,\,\,\,\,\,
    \left(
\begin{array}{ccc}
0 & e^{i(\alpha_{u} + \beta_{d})}\,\lambda_{uc^{\prime}} & e^{i(\alpha_{u} + \beta_{d})}\,\lambda_{ut^{\prime}}  \\
-e^{i(\alpha_{u} + \beta_{d})}\,\lambda_{uc^{\prime}}  &  0  & e^{i(2\beta_{d})}\,\lambda_{ct^{\prime}}     \\
-e^{i(\alpha_{u} + \beta_{d})}\,\lambda_{ut^{\prime}}  & -e^{i(2\beta_{d})}\,\lambda_{ct^{\prime}}   & 0 
\end{array}
\right)
  &\overset{!}{=}
  \left(
\begin{array}{ccc}
0 & \lambda_{uc^{\prime}} & \lambda_{ut^{\prime}}  \\
-\lambda_{uc^{\prime}}  &  0  & \lambda_{ct^{\prime}}     \\
-\lambda_{ut^{\prime}}  & -\lambda_{ct^{\prime}}   & 0 
\end{array}
\right)\\
\nonumber 
\\
 \label{eq:LQ1diquarkDDb}
\lambda_{dd}:&& \,\,\,\,\,\,\,\,\,\,
    \left(
\begin{array}{ccc}
0 & e^{i(\alpha_{d}+\beta_{d})}\,\lambda_{ds^{\prime}} & e^{i(\alpha_{d}+\beta_{d})}\,\lambda_{db^{\prime}}  \\
-e^{i(\alpha_{d}+\beta_{d})}\,\lambda_{ds^{\prime}}  &  0  & e^{i(2\beta_{d})}\,\lambda_{sb^{\prime}}     \\
-e^{i(\alpha_{d}+\beta_{d})}\,\lambda_{db^{\prime}}  & -e^{i(2\beta_{d})}\,\lambda_{sb^{\prime}}   & 0 
\end{array}
\right)
  &\overset{!}{=}
  \left(
\begin{array}{ccc}
0 & \lambda_{ds^{\prime}} & \lambda_{db^{\prime}}  \\
-\lambda_{ds^{\prime}}  &  0  & \lambda_{sb^{\prime}}     \\
-\lambda_{db^{\prime}}  & -\lambda_{sb^{\prime}}   & 0 
\end{array}
\right)\\
\nonumber 
\\
\label{eq:LQ1diquarkDUb}
\lambda_{du}:&& \,\,\,\,\,\,\,\,\,\,
    \left(
\begin{array}{ccc}
e^{i(\alpha_{d}+\alpha_{u})}\, \lambda_{du^{\prime}}  & e^{i(\alpha_{d}+\beta_{d})}\, \lambda_{dc^{\prime}} &  e^{i(\alpha_{d}+\beta_{d})}\,\lambda_{dt^{\prime}}  \\
e^{i(\beta_{d}+\alpha_{u})}\,\lambda_{su^{\prime}}  & e^{i(2\beta_{d})}\, \lambda_{sc^{\prime}}  & e^{i(2\beta_{d})}\, \lambda_{st^{\prime}}     \\
e^{i(\beta_{d}+\alpha_{u})}\,\lambda_{bu^{\prime}}  & e^{i(2\beta_{d})}\,\lambda_{bc^{\prime}}   &  e^{i(2\beta_{d})}\, \lambda_{bt^{\prime}}
\end{array}
\right)
  &\overset{!}{=}
  \left(
\begin{array}{ccc}
\lambda_{du^{\prime}} & \lambda_{dc^{\prime}} & \lambda_{dt^{\prime}}  \\
\lambda_{su^{\prime}}  &  \lambda_{sc^{\prime}}   & \lambda_{st^{\prime}}     \\
\lambda_{bu^{\prime}}  & \lambda_{bc^{\prime}}   &  \lambda_{bt^{\prime}}  
\end{array}
\right) \\
\nonumber 
\\
\label{eq:LQ1diquarkUDb}
\lambda_{ud}:&& \,\,\,\,\,\,\,\,\,\,
    \left(
\begin{array}{ccc}
e^{i(\alpha_{d}+\alpha_{u})}\, \lambda_{ud^{\prime}}  & e^{i(\alpha_{d}+\beta_{d})}\, \lambda_{cd^{\prime}} &  e^{i(\alpha_{d}+\beta_{d})}\,\lambda_{td^{\prime}}  \\
e^{i(\beta_{d}+\alpha_{u})}\,\lambda_{us^{\prime}}  & e^{i(2\beta_{d})}\, \lambda_{cs^{\prime}}  & e^{i(2\beta_{d})}\, \lambda_{ts^{\prime}}     \\
e^{i(\beta_{d}+\alpha_{u})}\,\lambda_{ub^{\prime}}  & e^{i(2\beta_{d})}\,\lambda_{cb^{\prime}}   &  e^{i(2\beta_{d})}\, \lambda_{tb^{\prime}}
\end{array}
\right)
  &\overset{!}{=}
  \left(
\begin{array}{ccc}
 \lambda_{ud^{\prime}}  & \lambda_{cd^{\prime}} &  \lambda_{td^{\prime}}  \\
\lambda_{us^{\prime}}  &  \lambda_{cs^{\prime}}  &  \lambda_{ts^{\prime}}     \\
\lambda_{ub^{\prime}}  & \lambda_{cb^{\prime}}  &  \lambda_{tb^{\prime}}
\end{array}
\right) 
\end{align}
For this particular solution, requiring $\alpha_{u} \neq - \beta_{d}$ is sufficient to set $\lambda_{uc^{\prime}} = \lambda_{ut^{\prime}} = 0$.  Similarly, from \eqref{eq:LQ1diquarkDDb},  $\alpha_{d} \neq - \beta_{d}$ implies $\lambda_{ds^{\prime}} = \lambda_{db^{\prime}} = 0$.  Additionally, both of these phase (in)equalities simultaneously kill all but the (1,1) element of the first rows and columns of \eqref{eq:LQ1diquarkDUb}-\eqref{eq:LQ1diquarkUDb}, and so the inequality $\alpha_{d} \neq - \alpha_{u}$ serves as the last condition necessary to completely forbid couplings with first-generation quarks.  Finally, we observe that maintaining $\beta_{d} \neq 0$ kills \emph{all} entries in the diquark Yukawa couplings.

\begin{table} [t]
\centering
\small
{\renewcommand{\arraystretch}{1.85}
\begin{tabular}{|c||c|c|}
\hline
\multicolumn{3}{|c|}{Additional Phase Equalities to Forbid Proton Decay}\\
\hline
\hline
 Class & Kill First-Generation Couplings & Kill All Diquark Couplings \\
\hline
A  & $\alpha_u \neq - \beta_d$, $\alpha_d \neq - \beta_d$, and $\alpha_d \neq - \alpha_u$ & $\beta_d \neq 0$ \\
\hline
B  & $\beta_d \neq - \beta_u$, $\alpha_d \neq - \beta_d$, $\alpha_d \neq - \beta_u$, and $\beta_d \neq 0$ & \checkmark \\
\hline
C  & $\beta_d \neq - \gamma_u$, $\alpha_d \neq - \beta_d$, $\alpha_d \neq - \gamma_u$, and $\beta_d \neq 0$ & \checkmark \\
\hline
\end{tabular}}
\caption{Additional phase relationships required to forbid proton decay in the models given in Table \ref{tab:finalphases}.  Equalities in the second column kill the dangerous couplings to first-generation quarks, and the third column shows what (if any) additional relationships are required to fully remove the diquark operators from the Lagrangian.}
\label{tab:protonphase}
\end{table}

Continuing, we organize the other patterns of Table \ref{tab:finalphases} into three classes, Class A matrices where $\beta_u = \gamma_u$ (like $\lambda_{QL}^{[e3A]}$), Class B matrices with $\alpha_u = \gamma_u$ (like $\lambda_{QL}^{[e3B]}$), and Class C matrices with $\alpha_u = \beta_u$ (like $\lambda_{QL}^{[e3C]}$).  For Class A matrices, the same relations derived for $\lambda_{QL}^{[e3A]}$ are sufficient to suppress proton decay, whereas with Class B matrices we instead observe from \eqref{eq:LQdiquarkUU} that two inequalities are necessary to nullify the first generation couplings in $\lambda_{uu}$: $\beta_d \neq - \beta_u$ and $\beta_d \neq 0$.  However, the first equality also sends the (2,3) block to zero, meaning that the entire coupling is set to zero, $\lambda_{uu} = 0$.  Of course, the relative difference in the phases of the up sector does not effect the $\lambda_{dd}$ coupling; \eqref{eq:LQ1diquarkDDb} holds when $\alpha_u = \gamma_u$, and the requirement that $\beta_d \neq 0$ from $\lambda_{uu}$ simultaneously sets the (2,3) block of this matrix to zero as well.  This still leaves the $\lambda_{ds^{\prime}}$ and $\lambda_{db^{\prime}}$ elements, which are forbidden if $\alpha_d \neq - \beta_d$, which forces the entire coupling matrix to again be null: $\lambda_{dd} = 0$.  Moving to $\lambda_{du}$, the only matrix element not killed by the combined phase constraints sourced from $\lambda_{uu}$ and $\lambda_{dd}$ is $\lambda_{dc^{\prime}}$, which is strictly null if $\alpha_d \neq - \beta_u$.  Requiring this inequality sets $\lambda_{du} = 0$, and the combined application of all these required phases automatically forces $\lambda_{ud} = 0$.  Finally, the relationships required for $\lambda_{QL}^{[e3C]}$ (the only Class C matrix we derived) mimic those of the Class B matrices; four relationships are required to kill couplings to first generation couplings, which when realized simultaneously nullify the entire set of couplings.

To conclude, proton decay can be forbidden via additional (mis)alignments of phases on top of those required in Table \ref{tab:finalphases}.  The additional relationships required for $\lambda_{QL}$ are determined by whether $\alpha_u = \gamma_u$, $\beta_u = \gamma_u$, or $\alpha_u = \beta_u$.  In the former case, three relationships kill all couplings to first generation quarks, and a fourth identically nullifies all diquark Yukawa couplings.  For the latter two cases, four relationships are required to kill couplings to first generation quarks, which simultaneously kills all other couplings as well.  These findings are summarized in Table \ref{tab:protonphase}.

\section{$SU(2)$ analysis for the vectors $\Delta_{(1,3)}^{\mu}$}
\label{sec:otherstatesV1V3}

Up to this point all of our considerations have been for the scalar triplet $\Delta_{3}$.  However, as is clear in \eqref{eq:LLyukFLAV}, the vector singlet and triplet leptoquarks $\Delta_{(1,3)}^{\mu}$ introduce quark-lepton couplings with slightly different normalizations and definitions than in the scalar triplet case.  Also, the vector singlet $\Delta_{1}^{\mu}$ only permits LH couplings in the $u-\nu$ and $d-l$ sectors; additional RH couplings sourced from $x_{1}^{RR,\overline{RR}}$ obviously cannot be related via SU(2) transformations to the LH $d-l$ term. As a result, the patterns obtained after a residual symmetry analysis of the Yukawa sector of the vector Lagrangians are slightly different than those found for the scalar triplet, and we now discuss those subtleties.

We start by generalizing the results of Section \ref{sec:isolation}-\ref{sec:threecolumn} to the $\Delta_{3}^{\mu}$ state. From \eqref{eq:LLyukFLAV} we see that the change in conjugation structure for the fermion fields yields slightly different normalizations between $\lambda_{d\nu,ul,u\nu}^{V_{3}}$ and the phenomenologically relevant $\lambda_{dl}$.   Now taking $\lambda_{dl} \equiv - (U_{d}^{\dagger}\, x_{3}^{LL}\, U_{l})$, the other coupling matrices are given by
\begin{equation}
\label{eq:LQrelationsV3}
\lambda_{d\nu}^{V_{3}} = - \sqrt{2}\,\lambda_{dl}\,U_{PMNS}, \,\,\,\,\,\,\,\,\,\,\,\, \lambda_{ul}^{V_{3}} = - \sqrt{2}\,U_{CKM} \,\lambda_{dl}, \,\,\,\,\,\,\,\,\,\,\,\, \lambda_{u\nu}^{V_{3}} = - U_{CKM} \,\lambda_{dl} \, U_{PMNS},
\end{equation}
which from \eqref{eq:LQrelations} we immediately read off that
\begin{equation}
\label{eq:V3S3relate}
\lambda_{d\nu}^{V_{3}} = -2 \, \lambda_{d\nu}, \,\,\,\,\,\,\,\,\,\,\lambda_{ul}^{V_{3}} = -2 \, U_{CKM}\,U_{CKM}^{T} \, \lambda_{ul}, \,\,\,\,\,\,\,\,\,\,\lambda_{u\nu}^{V_{3}} = U_{CKM}\,U_{CKM}^{T} \, \lambda_{u\nu},
\end{equation}
where of course the practical impact of the factors of $U_{CKM} U_{CKM}^{T}$ is simply to send $U_{CKM}^{\star}$ in \eqref{eq:LQrelations} to $U_{CKM}$ in \eqref{eq:LQrelationsV3}.  

Despite the differences implied by \eqref{eq:V3S3relate}, the vector triplet includes the same number of couplings between quarks and leptons as does the scalar triplet, and therefore the residual symmetry analysis proceeds analogously to that in Sections \ref{sec:isolation}-\ref{sec:threecolumn}.  In particular, the isolation solutions and the six two-columned solutions given in Table \ref{tab:finalphases}, up to \eqref{eq:V3S3relate}, are also found for $\Delta_{3}^{\mu}$, but they correspond to slightly different phase relations due to the conjugation of the quark states ($\bar{d}^{C \, i}_{L}$ vs. $\bar{d}^{i}_{L}$, e.g.), cf. eq.(\ref{eq:LLyukFLAV}); the phase relations for the vector SU(2) triplet thus appear modified by an overall minus sign in each of the quark phases relative to the scalar SU(2) triplet,
 \begin{equation}
\label{eq:LQoverconstrainV3}
    \left(
\begin{array}{ccc}
e^{i(-\alpha_{d}+\alpha_{l})}\,\lambda_{de} & e^{i(-\alpha_{d}+\beta_{l})}\, \lambda_{d\mu} &  e^{i(-\alpha_{d}+\gamma_{l})}\,\lambda_{d\tau}  \\
e^{i(-\beta_{d}+\alpha_{l})}\,\lambda_{se}  & e^{i(-\beta_{d}+\beta_{l})}\, \lambda_{s\mu}  & e^{i(-\beta_{d}+\gamma_{l})}\, \lambda_{s\tau}     \\
e^{i(-\gamma_{d}+\alpha_{l})}\,\lambda_{be}  & e^{i(-\gamma_{d}+\beta_{l})}\,\lambda_{b\mu}   &  e^{i(-\gamma_{d}+\gamma_{l})}\,\lambda_{b\tau}  
\end{array}
\right)
  \overset{!}{=}
  \left(
\begin{array}{ccc}
\lambda_{de} & \lambda_{d\mu} & \lambda_{d\tau}  \\
\lambda_{se}  & \lambda_{s\mu}  & \lambda_{s\tau}     \\
\lambda_{be}  & \lambda_{b\mu}   & \lambda_{b\tau}  
\end{array}
\right)\,,
\end{equation}
such that, upon performing all relevant manipulations, minus signs appear in the final phase equalities of Table \ref{tab:finalphases}.  For example, the solution leading to $\lambda_{QL}^{e\mu1A}$ that, for the scalar triplet, appears together with phase relation
$\lbrace$ $\beta_{d}$, $\gamma_{d}$, $-\beta_{\nu}$,  $-\gamma_{\nu}$,  $-\alpha_{l}$,  $-\beta_{l}$,  $\beta_{u}$,  $\gamma_{u}$ $\rbrace$, instead appears for the vector triplet with phase relation
$\lbrace$ $\beta_{d}$, $\gamma_{d}$, $\beta_{\nu}$,  $\gamma_{\nu}$,  $\alpha_{l}$,  $\beta_{l}$,  $\beta_{u}$,  $\gamma_{u}$ $\rbrace$ (and similarly for the other solutions).  All of these relations are also given in Table \ref{tab:finalphases}.

For the SU(2) singlet vector $\Delta_{1}^{\mu}$, we now define $\lambda_{dl} \equiv (U_{d}^{\dagger}\,x_{1}^{LL}\,U_{l})$ and from \eqref{eq:LLyukFLAV} obtain the following normalization for $\lambda_{u\nu}^{V_{1}}$:
\begin{equation}
\label{eq:LQrelationsV1}
\lambda_{u\nu}^{V_{1}} = U_{CKM} \, \lambda_{dl} \, U_{PMNS}\,\,\,\,\,\,\,\,\,\, \Longrightarrow \,\,\,\,\,\,\,\,\,\, \lambda_{u\nu}^{V_{1}} = - U_{CKM}\,U_{CKM}^{T} \, \lambda_{u\nu}\,.
\end{equation}
Although the $d-l$ and $u-\nu$ couplings are the only LH operators for us to consider in our analysis for the vector singlet, it (perhaps surprisingly) turns out that the corresponding final solutions for $\lambda_{dl}$ and $\lambda_{u\nu}$ also map directly to the solutions found for the scalar triplet, up to the difference in normalization given in \eqref{eq:LQrelationsV1}. This is because applying solely $\lambda_{u \nu}^{V_{1}} = U_{CKM}\,\lambda_{dl}\,U_{PMNS}$ is sufficient to fix the CKM and PMNS relations obtained previously. Let us illustrate this with the isolation patterns of \eqref{eq:yukeisolation}.  For $\lambda_{dl}^{[e]}$, one obtains
\begin{equation}
\label{eq:unuV1}
\lambda_{u\nu}^{[e],V_{1}} =
\left(
\begin{array}{ccc}
U_{11}  \left( V^{\star}_{ub} \lambda_{be} + V^{\star}_{us} \lambda_{se} \right) & U_{12}  \left( V^{\star}_{ub} \lambda_{be} + V^{\star}_{us} \lambda_{se} \right) & U_{13} \left( V^{\star}_{ub} \lambda_{be} + V^{\star}_{us} \lambda_{se} \right) \\
U_{11}  \left( V^{\star}_{cb} \lambda_{be} + V^{\star}_{cs} \lambda_{se} \right) & U_{12} \left( V^{\star}_{cb} \lambda_{be} + V^{\star}_{cs} \lambda_{se} \right)  & U_{13} \left( V^{\star}_{cb} \lambda_{be} + V^{\star}_{cs} \lambda_{se} \right) \\
U_{11}  \left( V^{\star}_{tb} \lambda_{be} + V^{\star}_{ts} \lambda_{se} \right) & U_{12} \left( V^{\star}_{tb} \lambda_{be} + V^{\star}_{ts} \lambda_{se} \right) & U_{13} \left( V^{\star}_{tb} \lambda_{be} + V^{\star}_{ts} \lambda_{se} \right)
\end{array}
\right)\,.
\end{equation}
Our symmetry constraints demand that one column be set to zero, and an equality analogous to \eqref{eq:ulconstraint} implies that this can only be fully achieved with the PMNS matrix elements, for which only $U_{13}$ provides a reasonable approximation to zero.  This kills the third column in \eqref{eq:unuV1}.  Yet we must also remove a row from $\lambda_{u\nu}^{[e],V_{1}}$, and from \eqref{eq:ulconstraint} we have already observed that experimental bounds are satisfied when any of the three rows are set to zero, which implies $\lambda_{se} = - \lambda_{be} \frac{V^{\star}_{ub}}{V^{\star}_{us}}$, $\lambda_{se} = - \lambda_{be} \frac{V^{\star}_{cb}}{V^{\star}_{cs}}$, or $\lambda_{se} = - \lambda_{be} \frac{V^{\star}_{tb}}{V^{\star}_{ts}}$.  Up to CKM conjugation, these are precisely the same matrix-element relationships derived for the scalar triplet patterns $\lambda_{QL}^{e3(A,B,C)}$, as seen in Table \ref{tab:finalphases}.  Furthermore, had we instead considered $\mu$- or $\tau$-isolation for $\lambda_{dl}$, we would again observe that symmetry constraints cannot be met, as no PMNS element is small enough to approximate zero in these cases, thereby removing a column of $\lambda_{u\nu}^{[e],V_{1}}$ as required.
 
 The same derivations, when carried out on two-columned $\lambda_{dl}$ matrices, yield similar conclusions -- the solutions found for the scalar triplet are again found for the vector singlet, up to normalizations and conjugations.  They can be obtained from those given explicitly in Appendix \ref{sec:A}.  Also, no three-columned patterns are allowed for either the $u-\nu$ nor $d-l$ couplings of $\Delta_{1}^{\mu}$ --- when a single non-zero element is isolated on a row, the zeros of the corresponding column ultimately enforce $\lambda_{d i } = 0$ (with $i = (e,\mu,\tau)$) in \eqref{eq:yukemt}, and therefore these matrices reduce to their two-columned cousins, as seen in Section \ref{sec:threecolumn}.

However, one subtlety does arise in the vector singlet case when studying two-columned $\mu-\tau$ matrices for $\lambda_{dl}$, namely an ambiguity as to how the residual symmetry constraints are realized with respect to texture zeros in the matrix elements.  As it turns out, this subtlety does not change our conclusions --- no new matrices are generated.  However, let us elaborate by considering the $\lambda_{u\nu}^{V_{1}}$ coupling \emph{after} we have utilized our symmetries to remove the first column:
 \begin{equation}
\label{eq:unuV1mt}
\lambda_{u\nu}^{[\mu \tau],V_{1}} =
\frac{1}{U_{21}}\left(
\begin{array}{ccc}
0 &\left(U_{21} U_{32} - U_{22} U _{31} \right) \left( V^{\star}_{ub} \lambda_{b\tau} + V^{\star}_{us} \lambda_{s\tau} \right)  & \left(U_{21} U_{33} - U_{23} U _{31}  \right) \left( V^{\star}_{ub} \lambda_{b\tau} + V^{\star}_{us} \lambda_{s\tau} \right) \\
0 & \left( U_{21} U_{32} - U_{22} U _{31}\right) \left( V^{\star}_{cb} \lambda_{b\tau} + V^{\star}_{cs} \lambda_{s\tau}\right)   & \left( U_{21} U_{33} - U_{23} U _{31}   \right) \left(  V^{\star}_{cb} \lambda_{b\tau} + V^{\star}_{cs} \lambda_{s\tau} \right)  \\
0 & \left( U_{21} U_{32} - U_{22} U _{31}\right) \left(  V^{\star}_{tb} \lambda_{b\tau} + V^{\star}_{ts} \lambda_{s\tau} \right)  &\left(U_{21} U_{33} - U_{23} U _{31}   \right) \left( V^{\star}_{tb} \lambda_{b\tau} + V^{\star}_{ts} \lambda_{s\tau} \right) 
\end{array}
\right)\,,
\end{equation}
where in order to fully satisfy our symmetry demands we must also zero one of the rows of \eqref{eq:unuV1mt}.  Up to now, the relationship that satisfied this type of constraint was unique, i.e. only one of the bracketed expressions was phenomenologically consistent with null.  However, in this case, both the PMNS and CKM brackets in the (1,2) and (2,2) elements of \eqref{eq:unuV1mt} can be set to zero, and therefore one must consider all such possibilities when implementing the symmetry.  Generically labelling each row as $ \lbrace  0, (PMNS_{1} ) (CKM), (PMNS_{2})(CKM) \rbrace$, there are three possible combinations:\footnote{Not considering the measured values of the matrix elements of \eqref{eq:unuV1mt}, a fourth possibility emerges: $(PMNS_{1}) = 0$ and $(PMNS_{2}) = 0$.  While experimentally excluded for the normal hierarchy of couplings we consider in this paper, it should be noted that new patterns emerge if such a combination is allowed in other scenarios.  In particular, it does not enforce an additional relationship between the second and third generation down-quark leptoquark Yukawa couplings, $\lambda_{s\tau, b\tau}$, which permits a two-parameter matrix.}
\begin{enumerate}
\item $(CKM) = 0$ in both elements
\item $(PMNS_{1}) = 0$ and $(CKM) = 0$
\item $(PMNS_{2}) = 0$ and $(CKM) = 0$
\end{enumerate}
Option $1$ simply reduces to one of the patterns already discussed above, $\lambda_{QL}^{\mu\tau1A}$ in this case. Furthermore, options $2$ and $3$ also represent special cases of the $\lambda_{QL}^{\mu\tau1A}$ pattern.  After all, as can be seen in \eqref{eq:mt1Adnu}, while the symmetry considerations of the scalar triplet were unambiguous about what relationships had to be enforced amongst couplings,\footnote{Recall that the scalar (and vector) triplet Lagrangians include more operators than the vector singlet, and therefore more relationships between the symmetries of the different fermion sectors get enforced.  In particular, the $u-l$ and $d-\nu$ operators do not share the ambiguity of \eqref{eq:unuV1mt}.} the same brackets of PMNS elements still appear in the matrix.  If nature realizes a special alignment amongst them causing those terms to disappear, then an additional column of zeros will appear in $\lambda_{u\nu}^{V_{1}}$ and the overall PMNS structure of $\lambda_{dl}$ will also reflect the equality.

In addition to this subtlety, all of the solutions for the vector singlet are of course associated to fewer phase relations than in the scalar triplet (or even vector triplet case). This is simply due to the reduced number of operators in \eqref{eq:LLyukFLAV}; given terms relating only LH up quarks with neutrinos and LH down quarks with charged leptons, we do not have equalities between (e.g.) the phases of up quarks and charged leptons.  For example, in the $\lambda_{QL}^{e3A}$ case explored above, 
the pattern $\lbrace \beta_{d}, \gamma_{d}, -\alpha_{\nu}, - \beta_{\nu}, - \alpha_{l}, \beta_u, \gamma_u \rbrace$ derived for the scalar triplet is replaced with $\lbrace \beta_{d}, \gamma_{d}, \alpha_{l} \rbrace$ $\lbrace \alpha_{\nu}, \beta_{\nu}, \beta_u, \gamma_u \rbrace$ for the vector singlet (and similarly for the other solutions). Once again, these equalities are catalogued in Table \ref{tab:finalphases}.

Finally, with respect to $\Delta_{1}^{\mu}$, the patterns of the RR Yukawa couplings can also be constrained by residual symmetries applying to the RH quarks and leptons, in a specific model. One possibility is that they can be made to vanish entirely by imposing residual symmetries without relations between the phases. In terms of our analysis, because the RH sector is unphysical in the SM, we can not relate the respective $d-l$ and $u-\nu$ sectors in a model independent way, so the mass basis Yukawa couplings corresponding to $x_1^{RR}$ would not be as constrained as those that arose from $x_1^{LL}$.  We therefore do not address the possibilities for these RH operators in this paper.

\section{A Comment on Reducing the Symmetry of the Lagrangian}
\label{sec:reduce}

It is clear that the power of our approach lies in the imposition of the flavour symmetries actioned by $T_{u,d,l,\nu}$ on all Yukawa couplings present in our Lagrangian, and the further assumption that each $T$ have at least two distinct eigenvalues, so that they can legitimately be considered flavour symmetries.  However, the highly restricted set of patterns we have derived would be enlarged were we to reduce the amount of symmetry present in one or more fermionic sectors.  For example, imagine that in one sector we do not insist that the residual symmetry distinguishes at least two species, meaning its action can effectively be represented as a global phase rotation through the family.\footnote{Of course, the global phase rotations of one or more SM Yukawa sectors may be truly accidental, and not the physical remnants of some higher theory.  In this instance it does not make sense to even construct a generator $T$ for the sector, unless it is trivially the identity (i.e., the phases are set to zero).  We do not consider this case here, and point out that if no symmetry remains in either the up or down sectors, one is again forced to find another mechanism to forbid proton decay.}   This scenario can be realized, e.g., if $\mathcal{G}_{F}$ breaks (via a scalar flavon $\phi$ obtaining its vev $\langle \phi \rangle$) leaving as residual symmetry a subgroup which can be represented with equal entries along the diagonal.\footnote{Note the vev that preserves such a residual generator is not an irreducible triplet of the group, as $T \langle \phi \rangle = \langle \phi \rangle$ can't be solved for $T = e^{i \alpha} \mathbb{I}_{\bf{3}}$ in general.}
Despite becoming progressively less interesting, this could happen in all or multiple fermion sectors.

While cataloguing all of the explicit matrices permitted when one or more symmetries $T_{u,d,l,\nu}$ are trivialized (phases all set to be equal) is outside the scope of this paper, it is instructive to explore the different types of patterns allowed when one does.  Of course the most extreme scenario is where all of the generators $T_{u,d,l,\nu}$ are trivialized, $ T_{u,d,l,\nu} = e^{i \alpha_{u,d,l,\nu}} \mathbb{I}_{\bf{3}}$, such that we can make no predictions --- either no leptoquark couplings exist (in any given sector) or all nine do.  Yet more options exist between this trivial case and that of our paper's analysis.  Hence, as our guiding principle has been to utilize $\mathcal{R}_{K^{(\star)}}$ data to first constrain $\lambda_{dl}$, let us then study symmetry reduction in the down and/or charged-lepton sectors.

Consider the case where only one generator is active in the $d-l$ sector (take the down quarks).  In this case the analogue to our core equality \eqref{eq:LQoverconstrain} becomes 
 \begin{equation}
\label{eq:reduce1}
   T_{l} = e^{i \alpha_{l}} \mathbb{I}_{\bf{3}}\,\,\,\,\, \Longrightarrow \,\,\,\,\, \left(
\begin{array}{ccc}
e^{i(\alpha_{d}+\alpha_{l})}\,\lambda_{de} & e^{i(\alpha_{d}+\alpha_{l})}\, \lambda_{d\mu} &  e^{i(\alpha_{d}+\alpha_{l})}\,\lambda_{d\tau}  \\
e^{i(\beta_{d}+\alpha_{l})}\,\lambda_{se}  & e^{i(\beta_{d}+\alpha_{l})}\, \lambda_{s\mu}  & e^{i(\beta_{d}+\alpha_{l})}\, \lambda_{s\tau}     \\
e^{i(\beta_{d}+\alpha_{l})}\,\lambda_{be}  & e^{i(\beta_{d}+\alpha_{l})}\,\lambda_{b\mu}   &  e^{i(\beta_{d}+\alpha_{l})}\,\lambda_{b\tau}  
\end{array}
\right)
  \overset{!}{=}
  \left(
\begin{array}{ccc}
\lambda_{de} & \lambda_{d\mu} & \lambda_{d\tau}  \\
\lambda_{se}  & \lambda_{s\mu}  & \lambda_{s\tau}     \\
\lambda_{be}  & \lambda_{b\mu}   & \lambda_{b\tau}  
\end{array}
\right)\,,
\end{equation} 
where we have already recalled that $ \alpha_d \neq \beta_d = \gamma_d$ is required in order to permit entries simultaneously in the $s$ and $b$ rows of a given column.  But we now observe that $\alpha_d = - \alpha_l$ and $\beta_d = - \alpha_l$ are the only two solutions giving non-zero entries, and the first is irrelevant for resolving $\mathcal{R}_{K^{(\star)}}$.  Hence, a matrix with three zeros on the first row is the only allowed pattern when the charged-lepton generator is trivial,
 \begin{equation}
\label{eq:reduce2}
\lambda_{dl} = 
 \left(
 \begin{array}{ccc}
0 & 0 & 0  \\
\lambda_{se}  & \lambda_{s\mu}  & \lambda_{s\tau}     \\
\lambda_{be}  & \lambda_{b\mu}   & \lambda_{b\tau}  
\end{array}
\right)\,,
\end{equation}
which is clearly not one of the allowed solutions from before.  Similarly, had we trivialized the down-quark generator, we would observe that any two rows can be saved without violating our assumptions, yielding the same patterns as in \eqref{eq:yukem}, but with three entries allowed per column:
\begin{equation}
\label{eq:yukem2}
\lambda_{dl} = 
\left(
\begin{array}{ccc}
\lambda_{de}   & \lambda_{d\mu} & 0  \\
\lambda_{se} & \lambda_{s\mu} & 0    \\
 \lambda_{be} & \lambda_{b\mu} & 0 
\end{array}
\right), \,\,\,\,\,\,
\lambda_{dl} =  
\left(
\begin{array}{ccc}
\lambda_{de}  & 0 & \lambda_{d\tau} \\
\lambda_{se} & 0 & \lambda_{s\tau}    \\
 \lambda_{be} & 0 & \lambda_{b\tau} 
\end{array}
\right), \,\,\,\,\,\,
\lambda_{dl} =  
\left(
\begin{array}{ccc}
0 &  \lambda_{d\mu}  & \lambda_{d\tau}  \\
0 &  \lambda_{s\mu}  & \lambda_{s\tau}    \\
0 & \lambda_{b\mu} & \lambda_{b\tau} 
\end{array}
\right)\,.
\end{equation}
Finally, the most extreme case of symmetry reduction in the $d-l$ operator would be to allow both the charged-lepton and down-quark generators to be trivialized, giving only two viable patterns for $\lambda_{dl}$:  all matrix elements allowed ($\alpha_{d} = - \alpha_{l}$) or no matrix elements allowed.

And yet we have said nothing about the impact of the full symmetry operations active in the up and neutrino sectors, which are still related to $\lambda_{dl}$ via SU(2) relations.    It is likely that the zeros enforced by these symmetries in other couplings reduces the number of free parameters in the matrices we have just derived, in precisely the same way they did when all symmetries were active.  For example, we have just shown that when a coupling is subject to one of two symmetry operators, there are still as many as three zeros that must be enforced after SU(2) rotation.  Furthermore, the $\lambda_{u\nu}$ coupling is still constrained to the original sets of allowed patterns, which could (in principle, not considering experimental bounds) enforce as many as seven zeros.  While it is beyond our scope to catalogue all such permutations, we simply emphasize that additional patterns for $\lambda_{dl}$ are allowed if the $T$ matrices are not active in all fermionic sectors, and these should still depend on fewer parametric degrees of freedom than in an environment free of flavour symmetries.

\section{Summary and Outlook}
\label{sec:conclude}

We have considered model-independent, flavour-symmetric leptoquark extensions of the SM in an effort to explain $\mathcal{R}_{K^{(\star)}}$  anomalies alongside of the SM flavour problem. In particular, we promote the natural phase freedoms of the SM Yukawa sector into Abelian `residual symmetries' with origins in a UV flavour theory that, perhaps upon being broken by flavon fields obtaining vevs aligned along specific orientations in flavour space, preserves these symmetries in each fermion mass sector. Our core assumption in this paper is that the same residual symmetries hold in the additional Yukawa terms that involve the leptoquark. While these symmetries are common by-products of complete models of flavour (including flavon models with leptoquarks, see e.g. \cite{Varzielas:2015iva}), their associated phenomenology can be studied without reference to specific model-building assumptions, e.g. the nature of the UV flavour symmetry, the number of flavon fields, the structure of their vacua, etc.  Our approach therefore describes a simplified model space.

Necessarily, in order to have non-degenerate fermion generations with non-trivial CKM and PMNS mixing, we conclude that the residual flavour symmetries act as diagonal phase matrices in the SM fermion mass basis.  Upon assuming that two generations of fermions are distinguished in each sector and that non-vanishing entries for $s$ and $b$ quarks exist in the novel leptoquark coupling down quarks to charged leptons (so that $\mathcal{R}_{K^{(\star)}}$ can be explained), the allowed patterns of matrices are severely restricted;  accounting for relevant precision flavour data under the assumption of a SM-like hierarchy of leptoquark couplings, we predict only six fully consistent models and three additional ones with $U^{13}_{PMNS} = 0$.  In all cases the leptoquark couplings depend on only one parametric degree of freedom, with matrix elements otherwise composed entirely of CKM and PMNS entries.  Interestingly, with one exception the resulting matrices are hierarchical, with entries in the first and second rows (corresponding to first and second generation quarks) always involving combinations of CKM elements that generate $\lambda ^3$ and $\lambda ^2$ suppressions respectively.  These results hold for all three leptoquarks we studied:  the scalar SU(2) triplet and vector SU(2) triplet and singlet.  Finally, proton decay is readily avoided with the same residual symmetry mechanism without the need for additional model building.

Due to the intense reduction of complex parameters in favor of known SM mixing elements, our results are extremely predictive and deserve further study.  It would be intriguing to perform an exhaustive phenomenological survey of different flavour observables sensitive to the new leptoquark couplings, both for the normal hierarchy considered here and its generalizations.  For example, hints of LNU also persist in $b \rightarrow c$ transitions as encoded in the ratio observable $\mathcal{R}_{D^{(*)}}$ \cite{Lees:2012xj,Lees:2013uzd,Huschle:2015rga,Aaij:2015yra}, for which our simplified models will give clear (and testable) BSM signals.  We plan to address these and other predictions in a future publication. 
In addition, we are also interested in exploring the UV origins of the specific Abelian residual symmetries implied by the phase relations presented in Table \ref{tab:finalphases}.  One could (e.g.) perform a bottom-up (and model-independent) scan of finite groups along the lines of \cite{Talbert:2014bda,Varzielas:2016zuo} in order to expose non-Abelian discrete symmetries closed by the active residual generators $T_{u,d,l,\nu}$, or one could attempt to build a complete UV model whose scalar sector realizes the special symmetry breaking embedded in our simplified models.

We also emphasize that our residual flavour symmetry approach represents a novel means of constraining generic leptoquark extensions of the SM, regardless of whether or not the $\mathcal{R}_{K^{(\star)}}$ anomalies withstand further experimental scrutiny.  Indeed, an abundance of Yukawa sector parameters and the need for additional modeling to prevent proton decay represent common theoretical nuisances that must be overcome in BSM leptoquark environments.  Both are naturally achieved in our framework.

\section*{Acknowledgements}
We are extremely grateful to Gudrun Hiller, who provided many key insights and support during the development of this work.  We also thank Jordan Bernigaud for many helpful discussions.  IdMV acknowledges funding from the Funda\c{c}\~{a}o para a Ci\^{e}ncia e a Tecnologia (FCT) through
the contract IF/00816/2015 and partial support by Funda\c{c}\~ao para a Ci\^encia e a Tecnologia
(FCT) through projects CFTP-FCT Unit 777 (UID/FIS/00777/2013), CERN/FIS-PAR/0004/2017 and PTDC/FIS-PAR/29436/2017 which are partially funded through POCTI (FEDER), COMPETE, QREN and EU.
J.T. acknowledges research and travel support from DESY, thanks Jure Zupan, Jared Evans, and Yuval Grossman for interesting discussions on the subject, and thanks Fady Bishara for inspiring the title of this work.

\appendix
\section{List of Explicit Yukawa Couplings \label{app:final}}
\label{sec:A}

In what follows we give the matrix representations for the patterns of leptoquark Yukawa couplings derived in Sections \ref{sec:isolation} - \ref{sec:otherstatesV1V3}, and referenced alongside of their corresponding phase equalities in Table \ref{tab:finalphases}.  In particular, we show the explicit results for the scalar triplet $\Delta_{3}$, and we recall the definition of the CKM matrix elements given in the text:
\begin{align}
(U^{ij}_{CKM})^{\star} = V_{ij}\,.
\end{align}
Alongside of the exact predictions, we also provide numerically approximate forms that may be easier to manipulate for phenomenology.  In generating these we have approximated the CKM matrix with a leading-power expansion in the Cabibbo parameter $\lambda$, and also used best-fit values for leptonic mixing angles and CP-violating phase as reported in \cite{Esteban:2016qun,NuFit} for PMNS elements, in the normal mass-ordering scenario for neutrinos.  Note that, because we only give central values for illustration, some elements may appear to violate the experimental bounds we used to derive the patterns in the first place.  However, our derivations were of course more conservative, as we considered the full error bands for CKM and PMNS elements as seen in \eqref{eq:num}. 

Finally, in order to obtain the respective patterns for the vector leptoquarks discussed in Section \ref{sec:otherstatesV1V3}, one notes \eqref{eq:LQrelationsV3}-\eqref{eq:V3S3relate} and \eqref{eq:LQrelationsV1} which indicate that the following simple procedure should be performed on the matrices of the scalar triplet:
\begin{enumerate}
\item Replace all $V_{ij}$ entries with $V_{ij}^*$.
\item $\lambda_{dl}^{V_{3}} \equiv \lambda_{dl}$
\item $\lambda_{d\nu}^{V_{3}} = -2 \, \lambda_{d\nu}$
\item $\lambda_{ul}^{V_{3}} = -2 \lambda_{ul}$
\item $\lambda_{u\nu}^{V_{3}} = \lambda_{u\nu}$
\end{enumerate}
for the vector triplet.  For the vector singlet, one instead applies:
\begin{enumerate}
\item Replace all $V_{ij}$ entries with $V_{ij}^*$.
\item $\lambda_{dl}^{V_{1}} \equiv \lambda_{dl}$
\item $\lambda_{u\nu}^{V_{1}} = -\lambda_{u\nu}$
\end{enumerate}
because the $d-\nu$ and $u-l$ couplings do not appear in its Lagrangian.

\subsection{Isolation Patterns}

\boxed{\bf{\lambda_{QL}^{e3A}}}
\\
\begin{align}
\lambda^{[e3A]}_{dl} &= \lambda_{be}
\left(
\begin{array}{ccc}
0  & 0 & 0  \\
-\frac{V_{ub}}{V_{us}} & 0 & 0    \\
1 & 0 & 0 
\end{array}
\right)
\simeq
\lambda_{be}
\left(
\begin{array}{ccc}
0  & 0 & 0  \\
-A \lambda ^2 (\rho + i \eta) & 0 & 0    \\
1 & 0 & 0 
\end{array}
\right) \\
\lambda_{d\nu}^{[e3A]} &=
\frac{\lambda_{be}}{\sqrt{2}} \left(
\begin{array}{ccc}
0 & 0 & 0 \\
-U_{11} \frac{V_{ub}}{V_{us}} & -U_{12} \frac{V_{ub}}{V_{us}}  & 0   \\
U_{11} & U_{12}  &  0 
\end{array}
\right)
\simeq
\lambda_{be} \left(
\begin{array}{ccc}
0 & 0 & 0 \\
-A \lambda ^2 (0.58) (\rho + i \eta) & -A \lambda ^2 (0.39)(\rho + i \eta) & 0 \\
0.58 & 0.39 &  0 
\end{array}
\right) \\
\lambda_{ul}^{[e3A]} &=
\frac{ \lambda_{be}}{\sqrt{2}} \left(
\begin{array}{ccc}
0  & 0 & 0 \\
\left(-\frac{V_{ub}V_{cs}}{V_{us}} + V_{cb} \right)  & 0 & 0  \\
\left(-\frac{V_{ub}V_{ts}}{V_{us}} + V_{tb} \right) & 0  & 0
\end{array}
\right)
\simeq
\lambda_{be} \left(
\begin{array}{ccc}
0  & 0 & 0 \\
A \lambda ^2 (0.71) (1-\rho - i \eta) & 0 & 0  \\
0.71 & 0  & 0
\end{array}
\right)  \displaybreak \\
\lambda_{u\nu}^{[e3A]} &=
\lambda_{be} \left(
\begin{array}{ccc}
0  & 0 & 0 \\
U_{11} \left(\frac{V_{ub}V_{cs}}{V_{us}} - V_{cb} \right) & U_{12} \left(\frac{V_{ub}V_{cs}}{V_{us}} - V_{cb} \right)   & 0 \\
U_{11} \left(\frac{V_{ub}V_{ts}}{V_{us}} - V_{tb} \right)  & U_{12} \left(\frac{V_{ub}V_{ts}}{V_{us}} - V_{tb} \right)  & 0
\end{array}
\right)
\\
&\simeq
\lambda_{be} \left(
\begin{array}{ccc}
0  & 0 & 0 \\
A \lambda ^2 (-0.82) (1 - \rho - i \eta) & A \lambda ^2 (-0.55) (1 - \rho - i \eta) & 0 \\
-0.82 & -0.55 & 0
\end{array}
\right)
\end{align}
\boxed{\bf{\lambda_{QL}^{e3B}}}
\\
\begin{align}
\lambda^{[e3B]}_{dl} &= \lambda_{be}
\left(
\begin{array}{ccc}
0  & 0 & 0  \\
-\frac{V_{cb}}{V_{cs}} & 0 & 0    \\
1 & 0 & 0 
\end{array}
\right)
\simeq
\lambda_{be}
\left(
\begin{array}{ccc}
0  & 0 & 0  \\
-A \lambda ^2 & 0 & 0    \\
1 & 0 & 0 
\end{array}
\right) \\
\lambda_{d\nu}^{[e3B]} &=
\frac{\lambda_{be}}{\sqrt{2}} \left(
\begin{array}{ccc}
0 & 0 & 0 \\
-U_{11} \frac{V_{cb}}{V_{cs}} & -U_{12} \frac{V_{cb}}{V_{cs}}  & 0   \\
U_{11} & U_{12}  &  0 
\end{array}
\right)
\simeq
\lambda_{be} \left(
\begin{array}{ccc}
0 & 0 & 0 \\
-A \lambda ^2 (0.58) & -A \lambda ^2 (0.39)& 0   \\
0.58 & 0.39 &  0 
\end{array}
\right) \\
\lambda_{ul}^{[e3B]} &=
\frac{ \lambda_{be}}{\sqrt{2}} \left(
\begin{array}{ccc}
\left(-\frac{V_{us}V_{cb}}{V_{cs}} + V_{ub} \right)   & 0 & 0 \\
0 & 0 & 0  \\
\left(-\frac{V_{cb}V_{ts}}{V_{cs}} + V_{tb} \right) & 0  & 0
\end{array}
\right)
\simeq
\lambda_{be} \left(
\begin{array}{ccc}
-A \lambda ^3 (0.71) (1 - \rho - i \eta) & 0 & 0 \\
0 & 0 & 0  \\
0.71 & 0  & 0
\end{array}
\right) \\
\lambda_{u\nu}^{[e3B]} &=
\lambda_{be} \left(
\begin{array}{ccc}
U_{11} \left(\frac{V_{us}V_{cb}}{V_{cs}} - V_{ub} \right)   & U_{12} \left(\frac{V_{us}V_{cb}}{V_{cs}} - V_{ub} \right)  & 0 \\
0 & 0  & 0 \\
U_{11} \left(\frac{V_{cb}V_{ts}}{V_{cs}} - V_{tb} \right)  & U_{12} \left(\frac{V_{cb}V_{ts}}{V_{cs}} - V_{tb} \right)  & 0
\end{array}
\right)
\\
&\simeq
\lambda_{be} \left(
\begin{array}{ccc}
A \lambda ^3 (0.82)  (1 - \rho - i \eta) & A \lambda ^3 (0.55) (1 - \rho - i \eta) & 0 \\
0 & 0  & 0 \\
-0.82 & -0.55 & 0
\end{array}
\right)
\end{align}

\boxed{\bf{\lambda_{QL}^{e3C}}}
\\
\begin{align}
\lambda^{[e3C]}_{dl} &= \lambda_{be}
\left(
\begin{array}{ccc}
0  & 0 & 0  \\
-\frac{V_{tb}}{V_{ts}} & 0 & 0    \\
1 & 0 & 0 
\end{array}
\right)
\simeq
\lambda_{be}
\left(
\begin{array}{ccc}
0  & 0 & 0  \\
\frac{1}{A \lambda ^2} & 0 & 0    \\
1 & 0 & 0 
\end{array}
\right) \\
\lambda_{d\nu}^{[e3C]} &=
\frac{\lambda_{be}}{\sqrt{2}} \left(
\begin{array}{ccc}
0 & 0 & 0 \\
-U_{11} \frac{V_{tb}}{V_{ts}} & -U_{12} \frac{V_{tb}}{V_{ts}}  & 0   \\
U_{11} & U_{12}  &  0 
\end{array}
\right)
\simeq
\lambda_{be}
\left(
\begin{array}{ccc}
0 & 0 & 0 \\
\frac{0.58}{A \lambda ^2} & \frac{0.39}{A \lambda ^2} & 0   \\
0.58 & 0.39 &  0 
\end{array}
\right) \\
\lambda_{ul}^{[e3C]} &=
\frac{ \lambda_{be}}{\sqrt{2}} \left(
\begin{array}{ccc}
\left(-\frac{V_{us}V_{tb}}{V_{ts}} + V_{ub} \right)   & 0 & 0 \\
\left(-\frac{V_{cs}V_{tb}}{V_{ts}} + V_{cb} \right) & 0  & 0 \\
0 & 0  & 0
\end{array}
\right)
\simeq
\lambda_{be}
\left(
\begin{array}{ccc}
\frac{0.71}{A \lambda} & 0 & 0 \\
\frac{0.71}{A \lambda ^2} & 0  & 0 \\
0 & 0  & 0
\end{array}
\right)  \\
\label{eq:unu32}
\lambda_{u\nu}^{[e3C]}  &=
\lambda_{be} \left(
\begin{array}{ccc}
U_{11} \left(\frac{V_{us}V_{tb}}{V_{ts}} - V_{ub} \right)   & U_{12} \left(\frac{V_{us}V_{tb}}{V_{ts}} - V_{ub} \right)  & 0 \\
U_{11}\left(\frac{V_{cs}V_{tb}}{V_{ts}} - V_{cb} \right)  & U_{12} \left(\frac{V_{cs}V_{tb}}{V_{ts}} - V_{cb} \right)     & 0   \\
0 & 0 & 0
\end{array}
\right) 
\simeq
\lambda_{be} \left(
\begin{array}{ccc}
-\frac{0.82}{A \lambda} & -\frac{0.55}{A \lambda} & 0 \\
-\frac{0.82}{A \lambda ^2} & -\frac{0.55}{A \lambda ^2} & 0   \\
0 & 0 & 0
\end{array}
\right)
\end{align}


\subsection{Two-columned Patterns (electron-muon)}


\boxed{\bf{\lambda_{QL}^{e\mu 1A}}}
\\
\begin{align}
\lambda^{[e\mu 1A]}_{dl} &= \lambda_{b\mu}
\left(
\begin{array}{ccc}
0  & 0 & 0  \\
\frac{V_{ub}}{V_{us}} \frac{U_{21}}{U_{11}} & -\frac{V_{ub}}{V_{us}} & 0 \\
 -\frac{U_{21}}{U_{11}} & 1 & 0 
\end{array}
\right)
\\
&\simeq
\lambda_{b\mu}
\left(
\begin{array}{ccc}
0  & 0 & 0  \\
-A \lambda ^2 (
   (0.39 -0.089 i) (\rho + i \eta ) & - A \lambda ^2 (\rho + i \eta ) & 0   \\
0.39 -0.089 i & 1 & 0 
\end{array}
\right) \\
\lambda_{d\nu}^{[e \mu 1A]} &=
\frac{\lambda_{b\mu}}{\sqrt{2}} \left(
\begin{array}{ccc}
0 & 0 & 0 \\
0 & \left(\frac{U_{12}U_{21}}{U_{11}} - U_{22} \right) (\frac{V_{ub}}{V_{us}})  & \left(\frac{U_{13}U_{21}}{U_{11}} - U_{23} \right) (\frac{V_{ub}}{V_{us}})    \\
0 & \left(-\frac{U_{12}U_{21}}{U_{11}} + U_{22} \right)  &  \left(-\frac{U_{13}U_{21}}{U_{11}} + U_{23} \right) 
\end{array}
\right)
\\
&\simeq
\lambda_{b\mu}
\left(
\begin{array}{ccc}
0  & 0 & 0  \\
0 & -A \lambda ^2 (0.58) (\rho +  i \eta ) & -A \lambda ^2 (0.50\, +0.039 i) (\rho + i \eta) \\
0 & 0.58 & 0.50\, +0.039 i
\end{array}
\right)  \\
\lambda_{ul}^{[e \mu 1A]} &=
\frac{ \lambda_{b\mu}}{\sqrt{2}} \left(
\begin{array}{ccc}
0  & 0 & 0 \\
\frac{U_{21}}{U_{11}}\left(\frac{V_{ub}V_{cs}}{V_{us}} - V_{cb} \right)  & \left(-\frac{V_{ub}V_{cs}}{V_{us}} + V_{cb} \right) & 0  \\
\frac{U_{21}}{U_{11}}\left(\frac{V_{ub}V_{ts}}{V_{us}} - V_{tb} \right)&  \left(-\frac{V_{ub}V_{ts}}{V_{us}} + V_{tb} \right)  & 0
\end{array}
\right)
\\
& \simeq
\lambda_{b\mu}
\left(
\begin{array}{ccc}
0  & 0 & 0  \\
A \lambda ^2 (0.28\, -0.063 i) (1-\rho-i\eta) & A \lambda ^2 (0.71) (1-\rho-i\eta) & 0   \\
0.28\, -0.063 i & 0.71 & 0 
\end{array}
\right) 
\\
\lambda_{u\nu}^{[e\mu 1A]} &=
\lambda_{b\mu} \left(
\begin{array}{ccc}
0  & 0 & 0 \\
0 & \left( \frac{U_{12} U_{21}}{U_{11}} - U_{22} \right) \left(-\frac{V_{ub}V_{cs}}{V_{us}} +V_{cb} \right)  &  \left( \frac{U_{13} U_{21}}{U_{11}} - U_{23} \right) \left(-\frac{V_{ub}V_{cs} }{V_{us}} +V_{cb} \right)  \\
0 & \left( \frac{U_{12} U_{21}}{U_{11}} - U_{22} \right) \left(-\frac{V_{ub}V_{ts}}{V_{us}} +V_{tb} \right) & \left( \frac{U_{13} U_{21}}{U_{11}} - U_{23} \right) \left(-\frac{V_{ub}V_{ts} }{V_{us}} +V_{tb} \right) 
\end{array}
\right)
\\
&\simeq
\lambda_{b\mu}
\left(
\begin{array}{ccc}
0  & 0 & 0  \\
0 & A \lambda ^2 (-0.82) (1 - \rho - i \eta) & A \lambda ^2 (-0.70-0.055 i) (1 - \rho - i \eta)   \\
0 & -0.82 & -0.70-0.055 i
\end{array}
\right)
\end{align}


\boxed{\bf{\lambda_{QL}^{e\mu 1B}}}
\\

\begin{align}
\lambda^{[e\mu 1B]}_{dl} &= \lambda_{b\mu}
\left(
\begin{array}{ccc}
0  & 0 & 0  \\
\frac{U_{21}}{U_{11}}\frac{V_{cb}}{V_{cs}}  & -\frac{V_{cb}}{V_{cs}} & 0    \\
 -\frac{U_{21}}{U_{11}} & 1 & 0 
\end{array}
\right)
\\
&\simeq
\lambda_{b\mu}
\left(
\begin{array}{ccc}
0  & 0 & 0  \\
- A \lambda ^2 (0.39-0.089 i) & - A \lambda ^2 & 0 \\
(0.39-0.089 i) & 1 & 0 
\end{array}
\right)
\\
\lambda_{d\nu}^{[e \mu 1B]} &=
\frac{\lambda_{b\mu}}{\sqrt{2}} \left(
\begin{array}{ccc}
0 & 0 & 0 \\
0 & \left(\frac{U_{12}U_{21}}{U_{11}} - U_{22} \right) (\frac{V_{cb}}{V_{cs}}) & \left(\frac{U_{13}U_{21}}{U_{11}} - U_{23} \right) (\frac{V_{cb}}{V_{cs}})    \\
0 & \left(-\frac{U_{12}U_{21}}{U_{11}} + U_{22} \right) &  \left(-\frac{U_{13}U_{21}}{U_{11}} + U_{23} \right)
\end{array}
\right)
\\
& \simeq
\lambda_{b\mu}
\left(
\begin{array}{ccc}
0  & 0 & 0  \\
0 & - A \lambda ^2 (0.58)  & - A \lambda ^2 (0.50 + 0.039 i) \\
0 & 0.58 & 0.50 + 0.039 i
\end{array}
\right) \displaybreak
\\
\lambda_{ul}^{[e \mu 1B]} &=
\frac{ \lambda_{b\mu}}{\sqrt{2}} \left(
\begin{array}{ccc}
\frac{U_{21}}{U_{11}}\left(\frac{V_{us}V_{cb}}{V_{cs}} - V_{ub} \right) & \left(-\frac{V_{us}V_{cb}}{V_{cs}} + V_{ub} \right) & 0  \\
0  & 0 & 0 \\
\frac{U_{21}}{U_{11}}\left(\frac{V_{cb}V_{ts}}{V_{cs}} - V_{tb} \right)  &  \left(-\frac{V_{cb}V_{ts}}{V_{cs}} + V_{tb} \right)  & 0
\end{array}
\right)
\\
&\simeq
\lambda_{b\mu}
\left(
\begin{array}{ccc}
- A \lambda^3 (0.28 - 0.063 i) (1 - \rho - i \eta)  & - A \lambda^3 (0.71) (1 - \rho - i \eta) & 0 \\
0 & 0 & 0 \\
0.28 - 0.063 i & 0.71  & 0
\end{array}
\right)
\\
\lambda_{u\nu}^{[e\mu 1B]} &=
\lambda_{b\mu} \left(
\begin{array}{ccc}
0 & \left( \frac{U_{12} U_{21}}{U_{11}} - U_{22} \right) \left(-\frac{V_{cb}V_{us}}{V_{cs}} + V_{ub} \right) &  \left( \frac{U_{13} U_{21}}{U_{11}} - U_{23} \right) \left(-\frac{V_{us}V_{cb}}{V_{cs}} + V_{ub} \right)  \\
0  & 0 & 0 \\
0 &\left(\frac{U_{12} U_{21} }{U_{11}} - U_{22} \right) \left(-\frac{V_{cb}V_{ts}}{V_{cs}} + V_{tb} \right) & \left( \frac{U_{13} U_{21}}{U_{11}} - U_{23} \right) \left(- \frac{V_{cb}V_{ts}}{V_{cs}} + V_{tb}  \right)\end{array}
\right)
\\
& \simeq
\lambda_{b\mu}
\left(
\begin{array}{ccc}
0  & - A \lambda^3 (-0.82) (1 - \rho - i \eta) & -A \lambda^3 (-0.70 - 0.055 i) (1 - \rho - i \eta) \\
0 & 0 & 0 \\
0 & -0.82 & -0.70-0.055 i 
\end{array}
\right)
\end{align}

\subsection{Two-columned Patterns (electron-tauon)}

\boxed{\bf{\lambda_{QL}^{e\tau 1A}}}
\\
\begin{align}
\lambda^{[e\tau 1A]}_{dl} &=
\lambda_{b\tau} \left(
\begin{array}{ccc}
0  & 0 & 0 \\
 \frac{U_{31}}{U_{11}} \frac{V_{ub}}{V_{us}} & 0 & -\frac{V_{ub}}{V_{us}}    \\
 -\frac{U_{31}}{U_{11}} & 0 & 1 
\end{array}
\right)
\\
&\simeq
\lambda_{b\tau}
\left(
\begin{array}{ccc}
0  & 0 & 0  \\
-A \lambda ^2 (-0.55-0.083 i) (\rho + i \eta) & 0 & -A \lambda ^2 (\rho + i \eta) \\
(-0.55-0.083 i) & 0 & 1 
\end{array}
\right)
\\
\lambda_{d\nu}^{[e \tau 1A]} &=
\frac{ \lambda_{b\tau}}{\sqrt{2}} \left(
\begin{array}{ccc}
0 & 0 & 0 \\
0 & \left(\frac{U_{12}U_{31}}{U_{11}} - U_{32} \right) (\frac{V_{ub}}{V_{us}})  & \left(\frac{U_{13}U_{31}}{U_{11}} - U_{33} \right) (\frac{V_{ub}}{V_{us}})  \\
0 & \left(-\frac{U_{12}U_{31}}{U_{11}} + U_{32} \right)  &  \left(-\frac{U_{13}U_{31}}{U_{11}} + U_{33} \right) 
\end{array}
\right)
\\
&\simeq
\lambda_{b\tau}
\left(
\begin{array}{ccc}
0  & 0 & 0  \\
0 & A \lambda ^2 (0.62)(\rho + i \eta)  & -A \lambda ^2 (0.52-0.042 i) (\rho + i \eta) \\
0 & -0.62 & (0.52-0.042 i)
\end{array}
\right)
\\
\lambda_{ul}^{[e \tau 1A]} &=
\frac{ \lambda_{b\tau}}{\sqrt{2}} \left(
\begin{array}{ccc}
0  & 0 & 0 \\
\frac{U_{31}}{U_{11}}\left(\frac{V_{ub}V_{cs}}{V_{us}} - V_{cb} \right) & 0 & \left(-\frac{V_{ub}V_{cs}}{V_{us}} + V_{cb} \right) \\
\frac{U_{31}}{U_{11}}\left(\frac{V_{ub}V_{ts}}{V_{us}} - V_{tb} \right)  & 0 &  \left(-\frac{V_{ub}V_{ts}}{V_{us}} + V_{tb} \right)
\end{array}
\right)
\\
&\simeq
\lambda_{b\tau}
\left(
\begin{array}{ccc}
0  & 0 & 0  \\
A \lambda ^2 (-0.39-0.058 i) (1 - \rho - i \eta) & 0 & A \lambda ^2 (0.71) (1 - \rho - i \eta) \\
(-0.39-0.058 i) & 0 & 0.71
\end{array}
\right)
\\
\lambda_{u\nu}^{[e\tau 1A]} &=
\lambda_{b\tau} \left(
\begin{array}{ccc}
0  & 0 & 0 \\
0 & \left( \frac{U_{12} U_{31}}{U_{11}} - U_{32} \right) \left(-\frac{V_{ub}V_{cs}}{V_{us}} +V_{cb} \right)  &  \left( \frac{U_{13} U_{31}}{U_{11}} - U_{33} \right) \left(-\frac{V_{ub}V_{cs} }{V_{us}} +V_{cb} \right)  \\
0 & \left( \frac{U_{12} U_{31}}{U_{11}} - U_{32} \right) \left(-\frac{V_{ub}V_{ts}}{V_{us}} +V_{tb} \right) & \left( \frac{U_{13} U_{31}}{U_{11}} - U_{33} \right) \left(-\frac{V_{ub}V_{ts} }{V_{us}} +V_{tb} \right) 
\end{array}
\right)
\\
&\simeq
\lambda_{b\tau}
\left(
\begin{array}{ccc}
0  & 0 & 0  \\
0 & A \lambda ^2 (0.88) (1 - \rho - i \eta) & A \lambda ^2 (-0.73+0.059 i) (1 - \rho - i \eta) \\
0 & 0.88 & (-0.73+0.059 i) 
\end{array}
\right)
\end{align}


\pagebreak
\boxed{\bf{\lambda_{QL}^{e\tau 1B}}}
\\
\begin{align}
\lambda^{[e\tau 1B]}_{dl} &=
\lambda_{b\tau} \left(
\begin{array}{ccc}
0  & 0 & 0 \\
 \frac{U_{31}}{U_{11}} \frac{V_{cb}}{V_{cs}} & 0 & -\frac{V_{cb}}{V_{cs}}    \\
  -\frac{U_{31}}{U_{11}} & 0 & 1 
\end{array}
\right)
\\
&\simeq
\lambda_{b\tau}
\left(
\begin{array}{ccc}
0  & 0 & 0  \\
-A \lambda ^2 (-0.55-0.083 i) & 0 & -A \lambda ^2 \\
(-0.55-0.083 i) & 0 & 1 
\end{array}
\right)
\\
\lambda_{d\nu}^{[e \tau 1B]} &=
\frac{\lambda_{b\tau}}{\sqrt{2}} \left(
\begin{array}{ccc}
0 & 0 & 0 \\
0 & \left(\frac{U_{12}U_{31}}{U_{11}} - U_{32} \right) (\frac{V_{cb}}{V_{cs}})  & \left(\frac{U_{13}U_{31}}{U_{11}} - U_{33} \right) (\frac{V_{cb}}{V_{cs}})    \\
0 & \left(-\frac{U_{12}U_{31}}{U_{11}} + U_{32} \right)  &  \left(-\frac{U_{13}U_{31}}{U_{11}} + U_{33} \right) 
\end{array}
\right)
\\
&\simeq
\lambda_{b\tau}
\left(
\begin{array}{ccc}
0 & 0 & 0  \\
0 & A \lambda ^2 (0.62)  & -A \lambda ^2 (0.52-0.042 i) \\
0 & -0.62 & (0.52-0.042 i)
\end{array}
\right)
\\
\lambda_{ul}^{[e \tau 1B]} &=
\frac{ \lambda_{b\tau}}{\sqrt{2}} \left(
\begin{array}{ccc}
\frac{U_{31}}{U_{11}}\left(\frac{V_{us}V_{cb}}{V_{cs}} - V_{ub} \right)& 0 & \left(-\frac{V_{us}V_{cb}}{V_{cs}} + V_{ub} \right)   \\
0  & 0 & 0 \\
\frac{U_{31}}{U_{11}}\left(\frac{V_{cb}V_{ts}}{V_{cs}} - V_{tb} \right) &  0 & \left(-\frac{V_{cb}V_{ts}}{V_{cs}} + V_{tb} \right) 
\end{array}
\right)
\\
&\simeq
\lambda_{b\tau}
\left(
\begin{array}{ccc}
-A \lambda ^3 (-0.39-0.058 i) (1 - \rho - i \eta) & 0 & -A \lambda ^3 (0.71) (1 - \rho - i \eta)\\
0 & 0 & 0 \\
(-0.39-0.058 i) & 0 & 0.71
\end{array}
\right) 
\\
\lambda_{u\nu}^{[e\tau 1B]} &=
\lambda_{b\tau} \left(
\begin{array}{ccc}
0 & \left( \frac{U_{12} U_{31}}{U_{11}} - U_{32} \right) \left(-\frac{V_{cb}V_{us}}{V_{cs}} + V_{ub} \right) &  \left( \frac{U_{13} U_{31}}{U_{11}} - U_{33} \right) \left(-\frac{V_{us}V_{cb}}{V_{cs}} + V_{ub} \right)  \\
0  & 0 & 0 \\
0 &\left(\frac{U_{12} U_{31} }{U_{11}} - U_{32} \right) \left(-\frac{V_{cb}V_{ts}}{V_{cs}} + V_{tb} \right) & \left( \frac{U_{13} U_{31}}{U_{11}} - U_{33} \right) \left(- \frac{V_{cb}V_{ts}}{V_{cs}} + V_{tb}  \right)\end{array}
\right)
\\
&\simeq
\lambda_{b\tau}
\left(
\begin{array}{ccc}
0 & -A \lambda ^3 (0.88) (1 - \rho - i \eta) & -A \lambda ^3 (-0.73+0.059 i) (1 - \rho - i \eta) \\
0 & 0 & 0 \\
0 & 0.88 & (-0.73+0.059 i)
\end{array}
\right)
\end{align}


\subsection{Two-columned Patterns (muon-tauon)}

\boxed{\bf{\lambda_{QL}^{\mu\tau 1A}}}
\\
\begin{align}
\lambda^{[\mu\tau 1A]}_{dl} &=
\lambda_{b\tau} \left(
\begin{array}{ccc}
0 & 0  & 0  \\
0 &  \frac{U_{31}}{U_{21}} \frac{V_{ub}}{V_{us}} & -\frac{V_{ub}}{V_{us}}    \\
0 & -\frac{U_{31}}{U_{21}} & 1 
\end{array}
\right)
\\
&\simeq
\lambda_{b\tau}
\left(
\begin{array}{ccc}
0  & 0 & 0  \\
0 &  - A \lambda^2 (1.3 +0.51 i) (\rho + i \eta) & - A \lambda^2 (\rho + i \eta) \\
0 & 1.3 +0.51 i & 1 
\end{array}
\right) \\
\label{eq:mt1Adnu}
\lambda^{[\mu\tau 1A]}_{d\nu} &=
\frac{\lambda_{b\tau}}{\sqrt{2}} \left(
\begin{array}{ccc}
0 & 0  & 0  \\
0 &  \left(\frac{U_{22}U_{31}}{U_{21}} - U_{32} \right) \frac{V_{ub}}{V_{us}} & \left(\frac{U_{23}U_{31}}{U_{21}} - U_{33} \right) \frac{V_{ub}}{V_{us}}    \\
0 & \left(-\frac{U_{22}U_{31}}{U_{21}} + U_{32} \right)  &   \left(-\frac{U_{23}U_{31}}{U_{21}} + U_{33} \right)   
\end{array}
\right)\\
&\simeq
\lambda_{b\tau}
\left(
\begin{array}{ccc}
0  & 0 & 0  \\
0 & - A \lambda^2 (0.12+0.29 i) (\rho + i \eta) & - A \lambda^2 (1.1+.26 i) (\rho + i \eta) \\
0 & 0.12+0.29 i & 1.1+.26 i
\end{array}
\right) \displaybreak
\\
\lambda^{[\mu\tau 1A]}_{ul} &=
\frac{\lambda_{b\tau}}{\sqrt{2}} \left(
\begin{array}{ccc}
0 & 0  & 0  \\
0 & \frac{U_{31}}{U_{21}} \left(\frac{V_{ub}V_{cs}}{V_{us}} - V_{cb} \right) & \left(-\frac{V_{ub}V_{cs}}{V_{us}} + V_{cb} \right)\\
0 &  \frac{U_{31}}{U_{21}} \left(\frac{V_{ub}V_{ts}}{V_{us}} - V_{tb} \right) & \left(-\frac{V_{ub}V_{ts}}{V_{us}} + V_{tb} \right)\end{array}
\right)\\
&\simeq
\lambda_{b\tau}
\left(
\begin{array}{ccc}
0  & 0 & 0  \\
0 & A \lambda^2 (0.92+0.36 i) (1 - \rho - i \eta) & A \lambda^2 (0.71) (1 - \rho - i \eta)\\
0 & 0.92+0.36 i &  0.71
\end{array}
\right) 
\\
\lambda^{[\mu\tau 1A]}_{u\nu} &=
\lambda_{b\tau} \left(
\begin{array}{ccc}
0 & 0  & 0  \\
0 &  \left(\frac{U_{22}U_{31}}{U_{21}} - U_{32}\right)\left(-\frac{V_{ub}V_{cs}}{V_{us}} + V_{cb}\right) & \left(\frac{U_{23}U_{31}}{U_{21}} - U_{33}\right)\left(-\frac{V_{ub}V_{cs}}{V_{us}} + V_{cb}\right)    \\
0 & \left(\frac{U_{22}U_{31}}{U_{21}} - U_{32}\right)\left(-\frac{V_{ub}V_{ts}}{V_{us}} + V_{tb}\right)   & \left(\frac{U_{23}U_{31}}{U_{21}} - U_{33}\right)\left(-\frac{V_{ub}V_{ts}}{V_{us}} + V_{tb}\right) \end{array}
\right)\\
&\simeq
\lambda_{b\tau}
\left(
\begin{array}{ccc}
0  & 0 & 0  \\
0 & A \lambda^2 (-0.18 - 0.41 i) (1 - \rho - i \eta) & A \lambda^2 (-1.6-0.37 i) (1 - \rho - i \eta)  \\
0 & -0.18 - 0.41 i & -1.6-0.37 i
\end{array}
\right)
\end{align}


\boxed{\bf{\lambda_{QL}^{\mu\tau 1B}}}
\\
\begin{align}
\lambda^{[\mu\tau 1B]}_{dl} &=
\lambda_{b\tau} \left(
\begin{array}{ccc}
0 & 0  & 0  \\
0 &  \frac{U_{31}}{U_{21}} \frac{V_{cb}}{V_{cs}} & -\frac{V_{cb}}{V_{cs}}    \\
0 & -\frac{U_{31}}{U_{21}} & 1 
\end{array}
\right)
\\
&\simeq
\lambda_{b\tau}
\left(
\begin{array}{ccc}
0  & 0 & 0  \\
0 & - A \lambda^2 (1.3+0.51 i) & - A \lambda^2 \\
0 & 1.3+0.51 i & 1 
\end{array}
\right)
\\
\lambda^{[\mu\tau 1B]}_{d\nu} &=
\frac{\lambda_{b\tau}}{\sqrt{2}} \left(
\begin{array}{ccc}
0 & 0  & 0  \\
0 &  \left(\frac{U_{22}U_{31}}{U_{21}} - U_{32} \right) \frac{V_{cb}}{V_{cs}} & \left(\frac{U_{23}U_{31}}{U_{21}} - U_{33} \right) \frac{V_{cb}}{V_{cs}}    \\
0 & \left(-\frac{U_{22}U_{31}}{U_{21}} + U_{32} \right)  &   \left(-\frac{U_{23}U_{31}}{U_{21}} + U_{33} \right)   
\end{array}
\right)
\\
& \simeq
\lambda_{b\tau}
\left(
\begin{array}{ccc}
0  & 0 & 0  \\
0 & -A \lambda^2 (0.12+0.29 i) & -A \lambda^2 (1.1 + 0.26 i)\\
0 & 0.12+0.29 i & 1.1 + 0.26 i
\end{array}
\right)
\\
\lambda^{[\mu\tau 1B]}_{ul} &=
\frac{\lambda_{b\tau}}{\sqrt{2}} \left(
\begin{array}{ccc}
0 &  \frac{U_{31}}{U_{21}} \left(\frac{V_{us}V_{cb}}{V_{cs}} - V_{ub} \right) & \left(-\frac{V_{us}V_{cb}}{V_{cs}} + V_{ub} \right) \\
0 & 0  & 0  \\
0 &  \frac{U_{31}}{U_{21}} \left(\frac{V_{cb}V_{ts}}{V_{cs}} - V_{tb} \right) & \left(-\frac{V_{cb}V_{ts}}{V_{cs}} + V_{tb} \right)\end{array}
\right)
\\
& \simeq
\lambda_{b\tau}
\left(
\begin{array}{ccc}
0  & - A \lambda^3 (0.92+0.36 i) (1 - \rho - i \eta) & - A \lambda^3 (0.71) (1 - \rho - i \eta) \\
0 & 0 & 0 \\
0 & 0.92+0.36 i &  0.71
\end{array}
\right)
\\
\lambda^{[\mu\tau 1B]}_{u\nu} &=
\lambda_{b\tau} \left(
\begin{array}{ccc}
0 &  \left(\frac{U_{22}U_{31}}{U_{21}} - U_{32}\right)\left(-\frac{V_{cb}V_{us}}{V_{cs}} + V_{ub}\right) & \left(\frac{U_{23}U_{31}}{U_{21}} - U_{33}\right)\left(-\frac{V_{us}V_{cb}}{V_{cs}} + V_{ub}\right)    \\
0 & 0  & 0  \\
0 & \left(\frac{U_{22}U_{31}}{U_{21}} - U_{32}\right)\left(-\frac{V_{cb}V_{ts}}{V_{cs}} + V_{tb}\right)   & \left(\frac{U_{23}U_{31}}{U_{21}} - U_{33}\right)\left(-\frac{V_{cb}V_{ts}}{V_{cs}} + V_{tb}\right) \end{array}
\right) 
\\
& \simeq
\lambda_{b\tau}
\left(
\begin{array}{ccc}
0  & A \lambda^3 (0.18 + 0.41 i) (1 - \rho - i \eta) & A \lambda^3 (1.6 + 0.37 i) (1 - \rho - i \eta) \\
0 & 0 & 0 \\
0 & -0.18-0.41 i & -1.6-0.37 i
\end{array}
\right)
\end{align}


\end{document}